\documentclass[10pt,a4paper]{emulateapj} 
\usepackage{lmodern}
\usepackage[latin1]{inputenc}
\usepackage[english]{babel}
\usepackage{amsmath}
\usepackage{amssymb}
\usepackage{latexsym}
\usepackage{epsfig}
\usepackage{subfigure}
\usepackage{natbib}
\usepackage{setspace}
\usepackage{longtable}
\usepackage{graphicx}
\usepackage{adjustbox}
\usepackage{changepage}
\usepackage{rotating}
\usepackage{blindtext}
\usepackage[11pt]{moresize}
\usepackage{hyperref}
\shorttitle{Galactic disk microlensing by
LSST}\shortauthors{Sajadian \& Poleski}
\begin{document}

\title{Prediction on detection and characterization of Galactic disk microlensing events by LSST}
\author{Sedighe Sajadian \altaffilmark{1,2}, Rados\l{}aw Poleski \altaffilmark{3}}
\altaffiltext{1}{Department~of~Physics,~Isfahan~University~of~Technology,~Isfahan~84156-83111,~Iran}\email{s.sajadian@cc.iut.ac.ir}

\altaffiltext{2}{CRANet-Isfahan,~Isfahan~University~of~Technology,~Isfahan,~84156-83111,~Iran}

\altaffiltext{3}{Department of Astronomy, Ohio State University, 140
W. 18th Ave., Columbus, OH 43210, USA}

\begin{abstract}
Upcoming LSST survey gives an unprecedented opportunity for studying
populations of intrinsically faint objects using microlensing
technique. Large field of view and aperture allow effective
time-series observations of many stars in Galactic disk and bulge.
Here, we combine Galactic models (for $|b|<10^{\rm{\circ}}$) and
simulations of LSST observations to study how different observing
strategies affect the number and properties of microlensing events
detected by LSST. We predict that LSST will mostly observe long
duration microlensing events due to the source stars with the
averaged magnitude around $22$ in $r-$band, rather than
high-magnification events due to fainter source stars. In Galactic
bulge fields, LSST should detect on the order of $400$ microlensing
events per square degree as compared to $15$ in disk fields.
Improving the cadence increases the number of detectable
microlensing events, e.g., improving the cadence from $6$ to $2$
days approximately doubles the number of microlensing events
throughout the Galaxy. According to the current LSST strategy, it
will observe some fields $900$ times during a $10-$year survey with
the average cadence of $\sim4\rm{-days}$ (I) and other fields
(mostly toward the Galactic disk) around $180$ times during a
$1-$year survey only with the average $\sim1\rm{-day}$ cadence (II).
We anticipate that the number of events corresponding to these
strategies are $7900$ and $34000$, respectively. Toward similar
lines of sight, LSST with the first observing strategy (I) will
detect more and on average longer microlensing events than those
observable with the second strategy. If LSST spends enough time
observing near Galactic plane, then the large number of microlensing
events will allow studying Galactic distribution of planets and
finding isolated black holes among wealth of other science cases.
\end{abstract}

\section{Introduction}\label{two}
The Large Synoptic Survey Telescope (LSST) is an optical and
wide-field telescope whose primary mirror has $8.4~\rm{m}$ diameter
and is now under construction in Chile \citep{lsstbook}. LSST is
supposed to observe the most of the visible sky every $4$ days. Its
field of view (FoV) and lifetime will be $9.6~\rm{deg^2}$ and $10$
years, respectively. The exposure time for each field will be
$30~\rm{sec}$. During a night, LSST will observe each field twice
with visits separated by $15-60$ minutes \citep{lsstbook2}. Hence,
LSST will take about $1000$ images for most fields in the visible
sky during its lifetime, i.e., the total number of visits in $10$
years is anticipated to be about $2.8$ million. 
Also $90$ percent of its time will be allocated to uniform
observations with a constant cadence and during the rest of time,
LSST will observe with other strategies.

LSST will (i) measure weak gravitational lensing to probe for any
signals of dark energy and dark matter, (ii) map our galaxy and its
objects, (iii) make an inventory of solar system components, and (iv)
detect transient phenomena in optical band such as supernovae
\citep{lsstbook}. However, other astrophysical events will be
detected during the LSST lifetime. For instance, if Galactic disk is
observed uniformly, then many microlensing events toward the
Galactic bulge and disk will be detected. In this regard,
\citet{Gould2013} studied the possibility of probing the planet
distribution in the Galactic plane by microlensing detection in
addition to the transit technique.

When the light of a background source star passes through the
gravitational field of a foreground massive object, it is bent
toward the center of gravity, the so-called gravitational lensing
event. In the Galactic scales, the light from a distant star can be
lensed and as a result magnified by a collinear massive object which
is called a gravitational microlensing event \citep{Einstein1936}.
In these events the light from the background source star is
magnified and produces two deformed images with unresolvable angular
separation.

\begin{figure*}
\centering
\subfigure[]{\includegraphics[angle=0,width=0.95\textwidth,clip=]{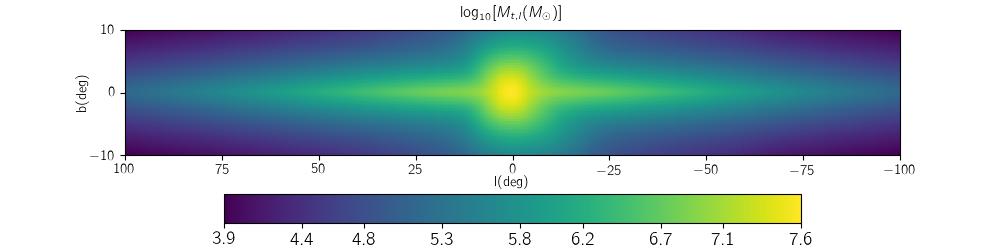}\label{fig0a}}
\subfigure[]{\includegraphics[angle=0,width=0.95\textwidth,clip=]{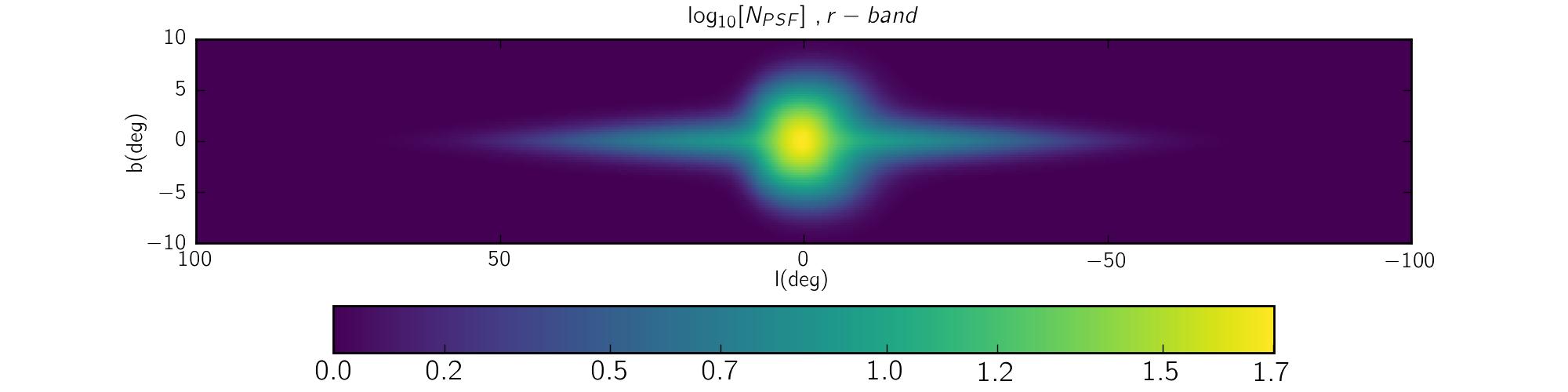}\label{fig0b}}
\caption{Figure \ref{fig0a}: Maps of the overall mass in the Galaxy $M_{t,l}\rm{[M_{\odot}]}$, which is the
integration over the cumulative mass densities due to all stellar
structures from the observer up to $20~\rm{kpc}$ toward each line of sight. Every line of sight covers the area
$\Omega_{l}=(0.25~\rm{deg})^{2}$. Figure \ref{fig0b}: the averaged
number of collinear stars whose light enters the LSST PSF of a
typical source star $N_{PSF}(l,b)$. The averaging is done over the
detectable microlensing events.}\label{fig0}
\end{figure*}
Detecting microlensing events toward the Galactic disk by a survey
telescope helps constraining the mass, spatial and velocity stellar
distributions in the Galactic thin and thick disks and the Galactic
bar. Probing microlensing events during $7$ seasons toward $4$
directions in the Galactic plane was done by EROS-II
\citep{Rahal2009}. As a result of this project, $27$ microlensing
candidates have been found. By performing a Monte Carlo simulation
according to the EROS-II observing strategy and comparing the
results from simulation with the real observation, \citet{Marc2017}
could test the Galactic models, different mass functions, etc. The
problems with microlensing events toward the Galactic plane are: (i)
we do not know source distances, whereas toward the Galactic bulge
almost all of the source stars belong to the Galactic bulge and are
located at the distance of approximately $8~\rm{kpc}$ far from us
and (ii) the interstellar extinction is high and variable in both
quantity and quality toward the Galactic disk. We lack detailed
knowledge about extinction in the disk because there is no group of
stars for which we apriority know colors and can be easily selected
(red clump stars serve in the bulge). The poorly known extinction
hampers optimizing observations and makes determination of event
properties harder, (i.e., the source angular size, the size of the
Einstein ring). These issues increase the number of degenerate
physical parameters. However, the overall distributions of the
observable parameters from observations can be used to examine the
different models in the Galaxy.

LSST is suppose to uniformly observe the galactic plane. As a
result, it will detect large number of microlensing events toward
the Galactic disk and bulge and its results will help to probe
different Galactic models of mass, velocity, special distribution,
etc. In this work, we aim to simulate LSST observations toward the
the Galactic bulge and disk with its observing strategies to predict
detections of the Galactic disk microlensing events. We aim to
specify the characteristics and statistics of the disk microlensing
events which will be detected during its era. However, we do not
consider the case of LSST providing supporting observations for the
WFIRST microlensing survey to measure free-floating planet masses.

We perform a Monte Carlo simulation of detectable microlensing
events during the LSST lifetime according to its observing strategy
in details explained in section (\ref{three}). In the following
section, we discuss the optical depth, the rate of events and the
number of observable events toward different directions resulted
from the simulation. Also we discuss the impact of observing cadence
on the results. In the section (\ref{five}), we simulate observing
microlensing events by considering two strategies for observation
(I) $900$ epochs during $10$ years observation with the
$3.9\rm{-day}$ cadence and (II) $180$ epochs during the first year
with the $0.9\rm{-day}$ cadence, to predict statistics and
properties of detectable microlensing events with these strategies.
We summarize the results and conclude in the section
(\ref{six}).\\

\section{Modeling of LSST microlensing observations}\label{three}
In order to simulate the microlensing events detectable by LSST, we
(i) simulate an ensemble of model microlensing events toward
different directions in the Galactic plane, (ii) generate synthetic
data points for each of them by assuming that these events are being
observed by LSST, and finally (iii) exert some criteria as
detectability threshold to simulated microlensing events to
determine (a) the LSST ability of detecting bulge and disk
microlensing events and (b) the characteristics of these events.
These three steps are discussed in the following subsections,
respectively.

\subsection{Simulating disk microlensing events}
In the following, we illustrate how to generate the parameters of
the source and the lens stars to make microlensing model light
curves. For the source stars, their locations are specified
according to the overall mass density throughout the Galaxy versus
distance in a given direction $dM/dD_{s}\propto\rho_{t}(l,b,D_{s})
D^{2}_{s} \Omega_{l}$, where $\rho_{t}(D_{s})$ is total mass density
due to all stellar structures in our galaxy, i.e., the thin and
thick disks, Galactic bulge (and bar), and the stellar halo, $D_{s}$
is the source distance from the observer, and $l$ and $b$ represent
the Galactic longitude and latitude respectively. We model these
mass densities using the Besan\c{c}on model
\citep{Robin2003,Robin2012}. One can find all details of these mass
density profiles in the Appendix B of \citet{Marc2017}. The map of
total mass $M_{t,l}=\int_{D_{s}} \rho_{t}(l,b,D_{s})
D_{s}^{2}~dD_{s}~\Omega_{l} $ in the unit of $M_{\odot}$ per line of
sight in the Galaxy is shown in Figure \ref{fig0a}. In the
simulation, the area corresponding to each line of sight is
$\Omega_{l}=(0.25~\rm{deg})^{2}$.

We indicate the intrinsic photometric properties of the source stars
in the same way as \citet{Penny2016} and use the old public version
of the Besan\c{c}on model\footnote{
We used the version available at \url{http://model.obs-besancon.fr/}.
There is a newer version from 2016: \url{http://modele2016.obs-besancon.fr/}}.
For each structure,
we simulate a sample of stars in CFHTLS-Megacam photometry system\footnote{
\url{http://www.cfht.hawaii.edu/Instruments/Imaging/Megacam/}}
without considering extinction. Then, the magnitudes in these
filters, $u^{\star}g'r'i'z'$, are converted to the magnitudes in the
Sloan Digital Sky Survey (SDSS) filters as \citep{Gwyn2008}:
\begin{eqnarray}
u^{\star}&=&u-0.241~(u-g),\nonumber\\
g'&=&g-0.153~(g-r),\nonumber\\
r'&=&r-0.024~(g-r),\nonumber\\
i'&=&i-0.085~(r-i),\nonumber\\
z'&=&z+0.074~(i-z);
\end{eqnarray}
LSST filters are the same as the SDSS photometry system
\citep{lsstbook,Fukugita1996}. These magnitudes can also be
transformed to the standard Johnson-Cousins photometry system\footnote{
\url{http://www.sdss3.org/dr8/algorithms/sdssUBVRITransform.php}}.
We do not simulate the magnitude of source stars in the $y-$band
filter. The Besan\c{c}on model does not give the magnitude of stars
in this band. However, the LSST photometric uncertainty in this band
is high, i.e., the 5$\sigma$ depth for point sources in $y-$band,
$m_{5,y}$, is $22.6$ whereas that for $z-$band is $24.45$.

In order to determine the apparent magnitude of the source stars, we
add the distance modulus and the extinction due to the interstellar
gas and dust to the absolute magnitudes. We use the 3D extinction
map presented by Marshal et al. (2006). They measured 3D
$K_{s}-$extinction map for the Galactic latitude in the range of
$b\in[-10:10^{\rm{\circ}}]$ and the Galactic longitude in the range
of $l\in [-100:100^{\rm{\circ}}]$ with the step $\Delta b=\Delta l=
0.25^\mathrm{\circ}$ and the distance step $0.5~\rm{kpc}$. Then, we
convert it to the extinction in $K-$band using $A_K=0.95~A_{K_s}$
\citep{Marshal2006}. The $K-$band extinction is converted to
the extinction in other bands using Cardelli et al.(1989)'s
relations. We assume $R_V$ of $2.5$ for Galactic bulge and $3.1$ for
thin disk, thick disk, and the stellar halo
\citep{Nataf2013,Cardelli1989}. We also add a Gaussian fluctuation
to each extinction value with the width in the range $[0.017,0.04]$
depending on the wavelength \citep{Cardelli1989}. The source radius
$R_{\star}$ is estimated using the Stefan-Boltezman relation:
$L=\sigma T^{4}_{eff} 4 \pi R^{2}_{\star}$, where $L$ is the source
luminosity, $T_{eff}$ is the effective surface temperature and
$\sigma$ is the Stefan-Boltzmann constant.

The blending effect in the LSST observations is significant, because
this telescope is anticipated to detect stars as faint as those with
$24.3~\rm{mag}$ in $r-$band. The number of such faint stars in the
Galaxy is very high, which makes high blending effects for these
faint stars. Therefore, it is crucial to consider and accurately
calculate the blending amount for each source star. In order to
calculate the contribution of background stars in the each LSST
Point Spread Function (PSF), we calculate the averaged number of
collinear stars toward the source line of sight whose light enters
the PSF area which is given by:
\begin{eqnarray}\label{nblend}
N_{PSF}(l,b)=\int_{D_{s}}~n_{t}(l,b,D_{s})~dD_{s}~D_{s}^{2}~\Omega_{PSF},
\end{eqnarray}
where $\Omega_{PSF}= \pi (\mathrm{FWHM}/2)^2$ in square arcsec is
the LSST PSF area for a typical source star. $\mathrm{FWHM}$ is the
Full Width at the Half Maximum of the brightness profile due to a
typical source star which is $(1.22,1.10,0.99,0.97,0.95)$ in arcsec
unit corresponding to the filters $ugriz$, respectively. The
$n_{t}(l,b,D_{s})$ is the total number density of stars in a given
direction and at the distance $D_{s}$ from the observer, which is
given by:
\begin{eqnarray}\label{nt}
n_{t}(l,b,D_{s})=\sum_{i=1}^{4}~\frac{\rho_{i}(l,b,D_{s})}{<M_{i}>}
\end{eqnarray}
where $\rho_{i}(l,b,D_{s})$ is the mass density for the $i$th
structure of our galaxy, $<M_{i}>$ is the averaged mass amount for
the $i$th structure and the summation is done over different
structures. The equation (\ref{nblend}) gives the average number of
blending stars and the true number in each measurement has a
gaussian fluctuation around this value with the width
$\sqrt{N_{\rm{PSF}}}$. Figure \ref{fig0b} represents the map of
$N_{\rm{PSF}}(l,b)$ in the Galaxy averaged over $N_{\rm{PSF}}$ of
observable microlensing events. For each of these blending stars, we
calculate their apparent magnitudes as explained in the previous
paragraph and as a result the overall flux due to all of blending
stars, i.e., $F_{base}= \sum_{i=1}^{N_{PSF}} 10^{-0.4~m_{i}}$. One
of these blending stars is the source star itself. The blending
parameter is given by $f_{b}=F_{\star}/F_{base}$, where $F_{\star}=
10^{-0.4m_{\star}}$ is the flux of the source star. In the real
observation, if the overall magnitude due to all blending stars into
a PSF reaches to the detection threshold, in that PSF one star is
discerned, whereas in the simulation all of these blending stars
pass the detectability criterion. In order to accordingly correct
the simulation, we weight each simulated source star with its
blending factor $f_{b}$.

For the lens population, we first indicate the lens distance from
the observer $D_{l}$ using the lensing probability function
$d\Gamma/dD_{l} \propto
\rho_{t}(D_{l})\sqrt{D_{l}(D_{s}-D_{l})/D_{s}}$. According to the
contribution of different structures in the total mass density at
the location of the lens $\rho_{t}(D_{l})$, we determine to which
structure the lens belongs to. Then, the lens mass is chosen from the
corresponding mass density function to the lens structure. The
Besan\c{c}on model provides separate mass functions for different
structures. We assume both source and lens stars have global and
dispersion velocities \citep{Kayser1986,Binney2008}. In order to
determine the lens-source relative velocity, we (i) obtain the
velocity components in the observer coordinate system which axes are
parallel with and normal to the line of sight direction and (ii)
after projecting the source velocity in the lens plane subtract each
component of the lens and source velocities. In the Besan\c{c}on
model, the values of different components of dispersion velocities
depend on the stellar age and structure. In the simulation, we
discard the events with the Einstein crossing time shorter than
$0.5$ or longer than $300$ days. The longer microlensing events are
hard distinguish in real observations \citep[see, e.g.,][]{Mroz2017,Wyrzykowski2015b}.
The lens impact parameter is
chosen uniformly from the range $[0:1]$.

\begin{figure}
\subfigure[]{\includegraphics[angle=0,width=8.cm,clip=]{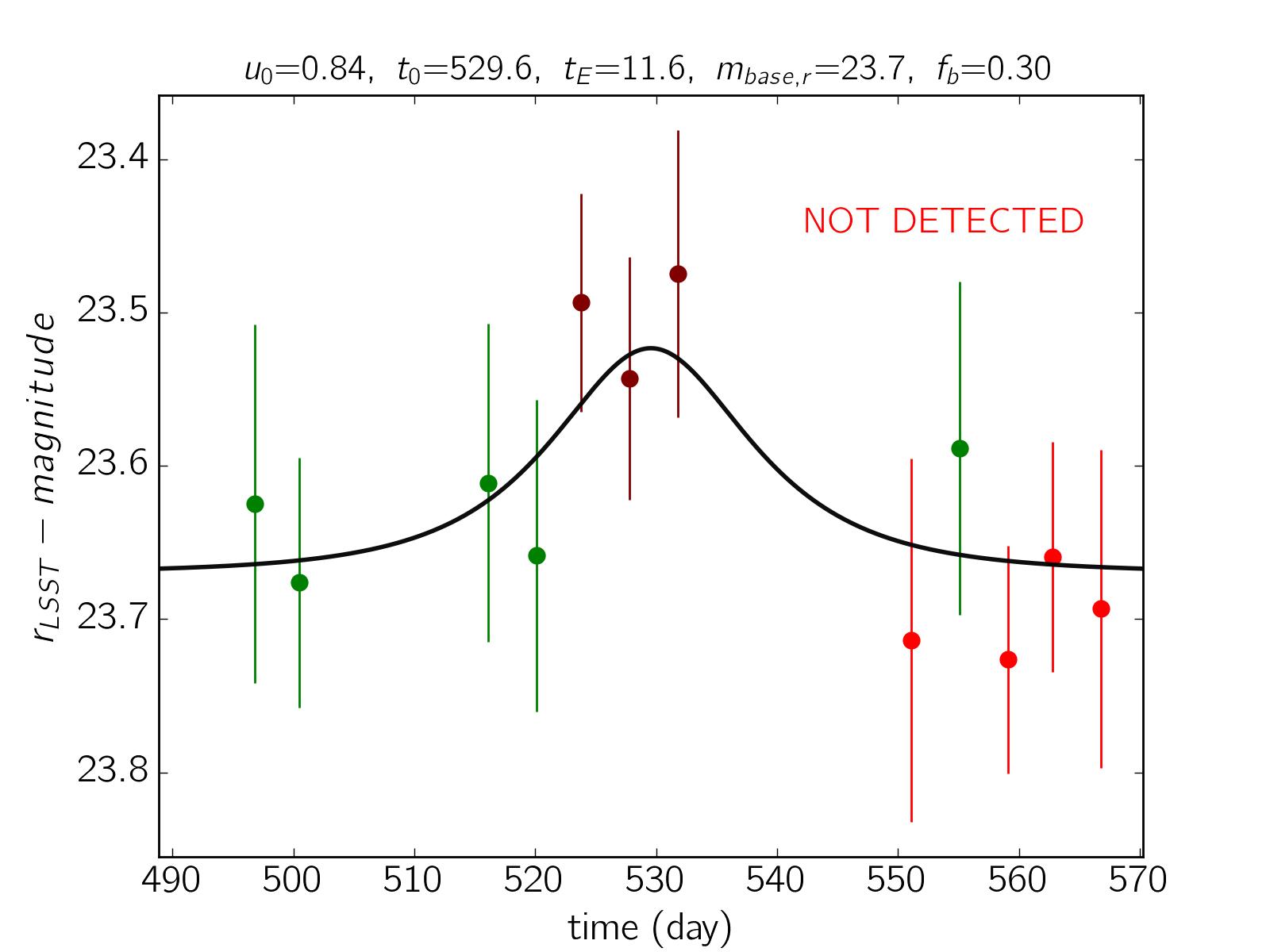}\label{fig1a}}
\subfigure[]{\includegraphics[angle=0,width=8.cm,clip=]{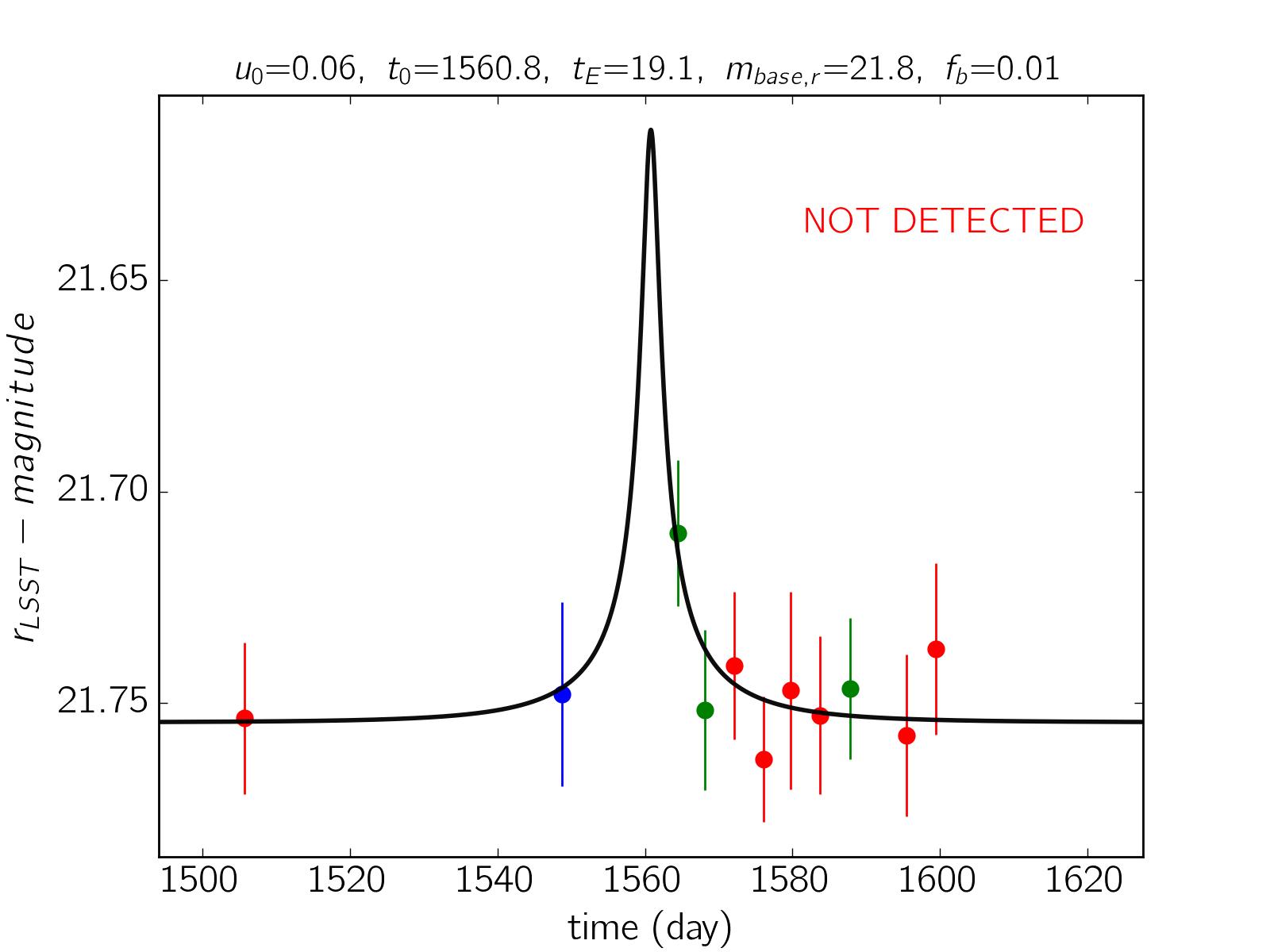}\label{fig1b}}
\subfigure[]{\includegraphics[angle=0,width=8.cm,clip=]{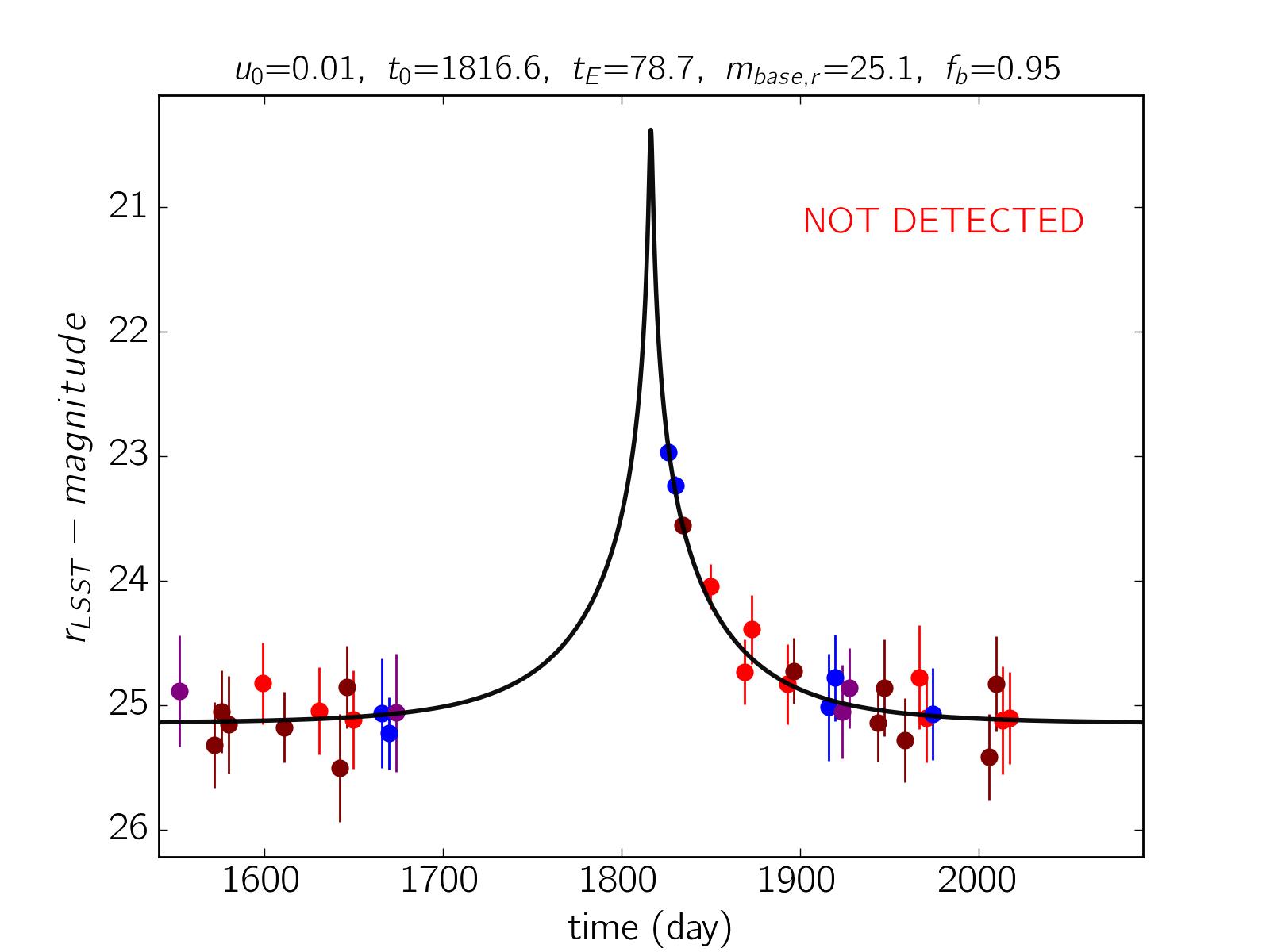}\label{fig1c}}
\caption{Example simulated microlensing events with the synthetic
data points taken by LSST which are not distinguishable. The
parameters used to make each of them are mentioned at the top of
each light curve. The color of data points indicates the filter
used: purple, blue, green, red and brown data points are taken
$ugriz$ filters respectively. The amounts of the background and
intrinsic source flux depends on the filter used. We assume that
these fluxes for each source star in all filters are measurable
during the LSST observation. As a result, we can convert the
measured magnitude in each filter to the corresponding magnitude in
$r-$band and easily model all data points with one light
curve.}\label{fig1}
\end{figure}
\subsection{Generating synthetic data points}
After generating the model microlensing events, the observability of
their source stars by LSST is checked. From simulated model
microlensing events, we ignore the microlensing events (i) which
source stars are too faint to be detected at least in one of the
LSST filters even when they are at the peak of their light curves
and (ii) which source stars are fainter than the saturation limit of
LSST in all filters. These events certainly are not observable.
We had the detection threshold (i.e., limiting magnitude) for
different LSST filters as given by \citet{lsstbook} (Table 2).

In order to test the observability of microlensing signatures of the
remaining microlensing events, we generate the hypothetically data
points taken by LSST and then verify if they can be discerned as
microlensing events by exerting some criteria. The data points are
simulated in the range $[t_{0}-3.5t_{\rm{E}}: t_{0}+3.5t_{\rm{E}}]$,
where $t_{\rm{E}}$ is the Einstein crossing time, $t_{0}$ is the
time of the closest approach that was uniformly chosen in the range
of $[0:T_{obs}]$, where $T_{obs}=10~\rm{yrs}$ is the LSST lifetime.
If the start time $t_{min}$ is less than zero (i.e.,
$t_{min}=t_{0}-3.5t_{\rm{E}}<0$) we start simulating data points
from $t_{min}=0.0$ and if the end time $t_{max}=t_{0}+3.5t_{\rm{E}}$
is greater than $T_{obs}$ we interrupt simulating data points on
$T_{obs}$. The exposure time is fixed on $30~\mathrm{seconds}$. For
timing the data points we also consider the seasonal gaps, i.e., the
LSST observations happen only during about seven continuous months
of each year. We assume that the weather is not suitable for
the LSST observation with the probability of $20\%$ for each night.
However, during each night, the uniform observation by LSST will
take place with the probability of $90\%$ which is considered in the
simulation.

For calculating the magnification factor, we consider the finite
size effect of the source star and use the adaptive contouring
algorithm \citep{Dominik2007} to calculate the magnification while
the lens distance from the source center is in the order of the
source radius projected on the lens plane. The synthetic data points
which are between saturation and detection limits of the LSST are
considered. We ignore the microlensing parallax effect during
magnification calculations.
\begin{figure}
\subfigure[]{\includegraphics[angle=0,width=8.cm,clip=]{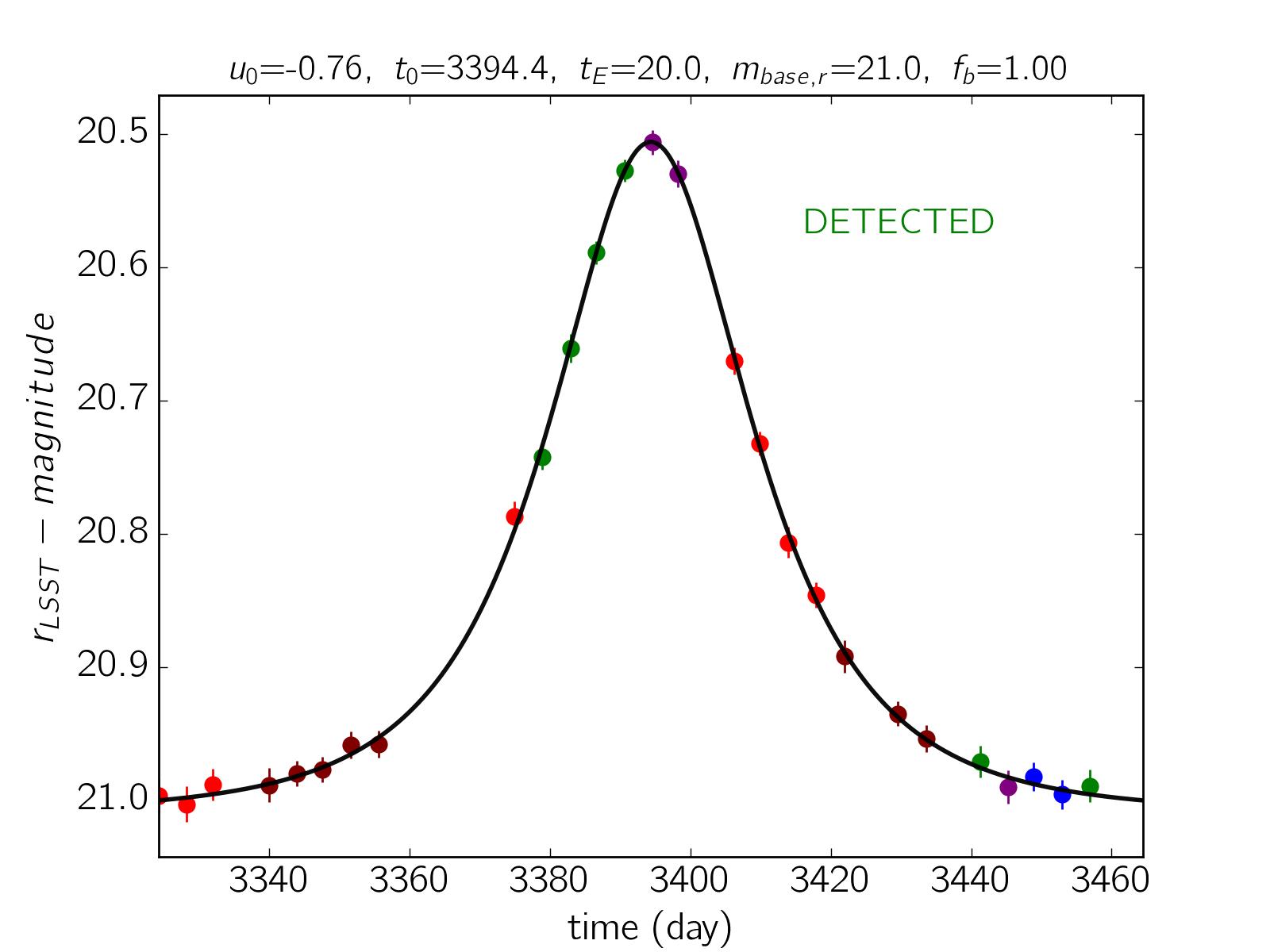}\label{fig1d}}
\subfigure[]{\includegraphics[angle=0,width=8.cm,clip=]{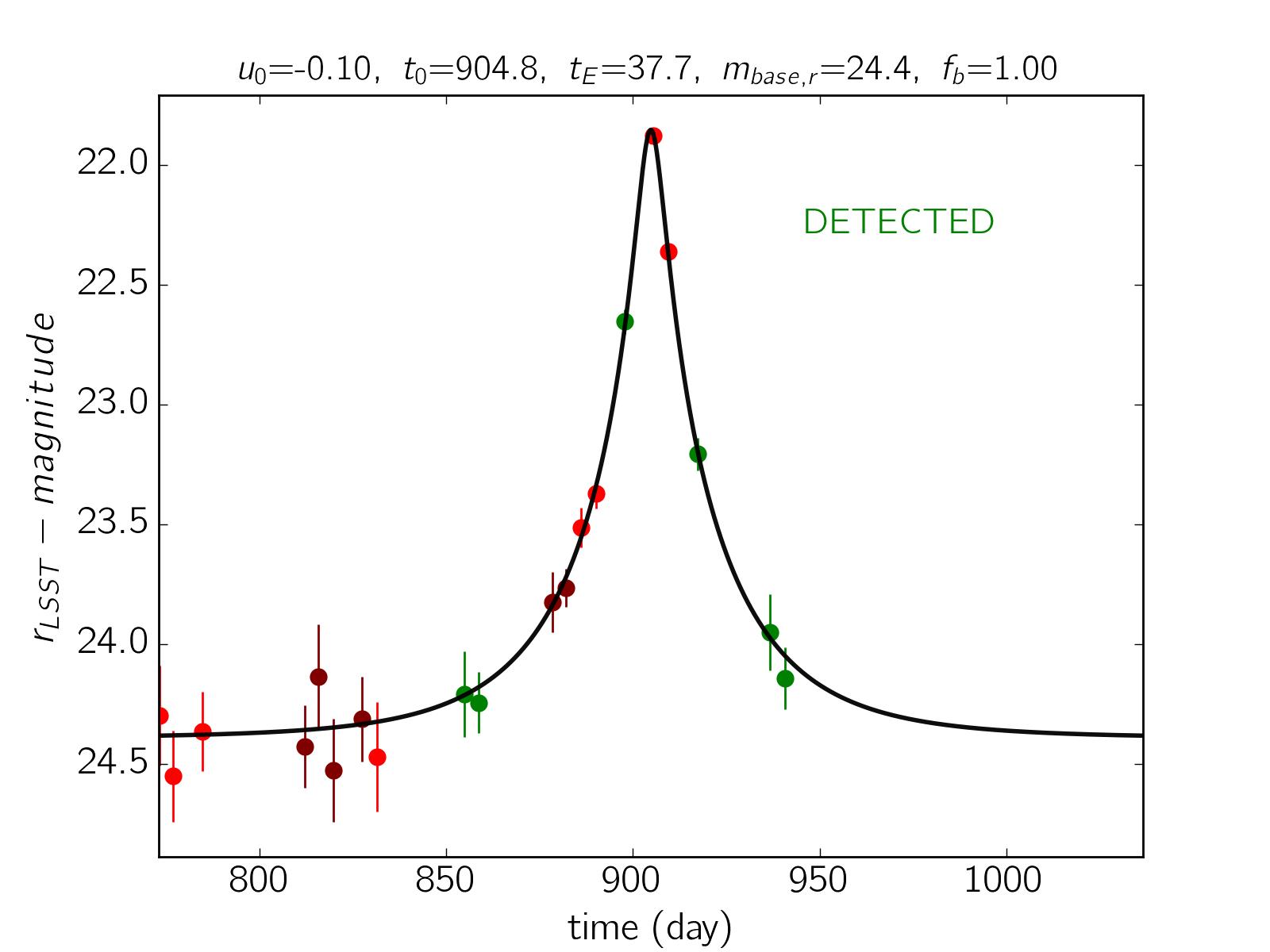}\label{fig1e}}
\subfigure[]{\includegraphics[angle=0,width=8.cm,clip=]{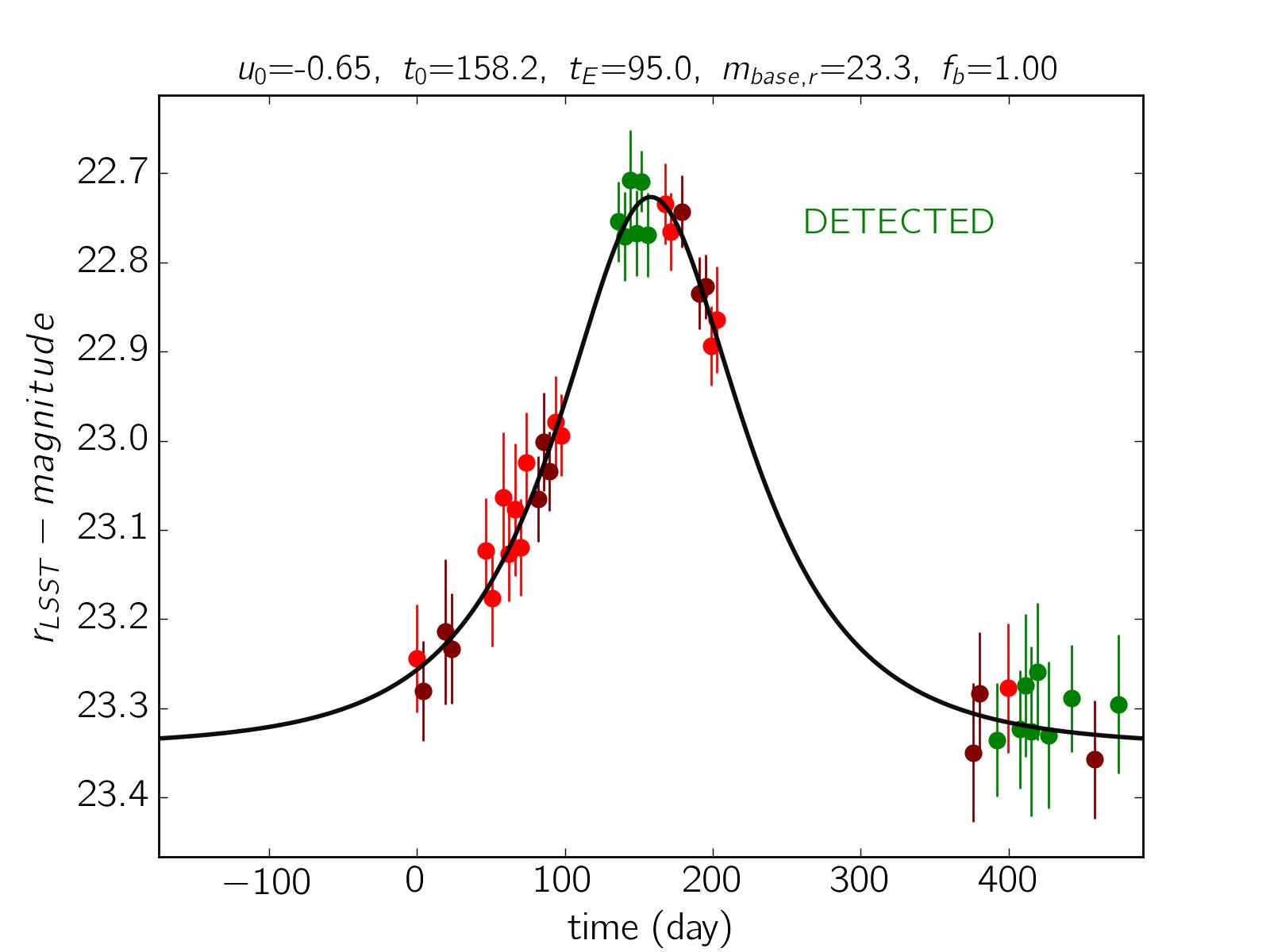}\label{fig1f}}
\caption{Example simulated microlensing events which are detectable
with LSST with the synthetic data points. More details can be found
in the caption of Figure (\ref{fig1}).}\label{fig10}
\end{figure}


We assume that the observed magnitude of each data point has a
gaussian fluctuation with respect to model value with the width
equal to expected photometric error filter $j$ ($j \in ugriz$):
$\sigma_{m,j}=\sqrt{\sigma^{2}_{sys}+\sigma^{2}_{rand,j}}$.
The $\sigma_{m,j}$ contains the random and systematic photometric
errors. We take into account these two uncertainties from
\citet{lsstbook}. For $\sigma_{rand}$ calculation we assumed sky
brightness as given by Table 2 of the mentioned paper, i.e.,
ignoring impact of other stars and actual zenith distance. The
photometric errors depend on wavelength and significantly decrease
for the bright stars. In order to determine the epoch, airmass and
filter used for taking each data point, we make a big ensemble of
these parameters from Operations Simulator (OpSim) with the approved
reference run minion$_{-}1016$\footnote{
\url{https://www.lsst.org/scientists/simulations/opsim}}. Each
data point is reported in one filter.

\subsection{Detectability criteria}\label{ceriteria}
After generating synthetic data points of each model microlensing
event, we probe if that event can be discerned. Our criteria for
detectability are (i) there are at least $4$ consecutive data points
(in any filter) with the observed magnification deviating from
constant flux more than $5~\sigma_{m,j}$, (ii) the difference of
$\chi^{2}$ from fitting the constant flux model and the microlensing
model to the synthetic data points is larger than $200$, and finally
(iii) the peak of the light curve should be between $t_{min}$ and
$t_{max}$. As a result from the third criterion, we ignore the
microlensing events for which the time of the closest approach is
out of the LSST lifetime.

Six example simulated microlensing events with the synthetic data
points are represented in Figures (\ref{fig1}) and (\ref{fig10}).
The parameters used to make each of them are mentioned at the top of
each light curve. $m_{base,r}$ is the baseline $r-$band brightness due to
cumulative fluxes entered in the source PSF. The $t_{\rm{E}}$ and
$t_{0}$ are expressed in the days. The amounts of the background and
intrinsic source flux depend on the filter used. We assume that
these fluxes for each source star in all filters are measurable
during the LSST observation. As a result, we can convert the
measured magnitude in each filter to the corresponding magnitude in
$r-$band and easily model all data points with one light curve,
as shown in the figures. 

The first microlensing event shown in Figure \ref{fig1a} is not
detectable by LSST, because of high blending. The source star is
magnified enough, but the blending causes the enhancement in the
stellar brightness due to the lensing effect to shrink and become on
the order of the photometric noise. The microlensing event shown in
Figure \ref{fig1b} is not recognizable although it is a
high-magnification event. Indeed, its time scale is too short in
comparison with the LSST cadence so that only a single data point is
taken during the magnification. The next event represented in Figure
\ref{fig1c} is not detectable also, although there is no blending
effect and the event's time scale is long enough. Because of the
seasonal gap, the peak of the light curve is not covered by the LSST
data points.

The microlensing events shown in Figure (\ref{fig10}) are observable
by LSST. The first event has the large impact parameter, i.e., small
lensing effect. Most microlensing events with small blending effect
and the bright source stars can be detected. The next one
(represented in Figure \ref{fig1e}) is detected. In contrast to the
high-magnification microlensing event \ref{fig1b} which was not
detectable, this high-magnification event is detected. The duration
of this event is long enough to take several data points while the
source star is magnified. The last microlensing event shown in
Figure \ref{fig1f} is also recognizable. The long-duration
microlensing events have higher chance to be detected during the
LSST lifetime in comparison to the short duration events.

\begin{figure*}
\centering
\subfigure[]{\includegraphics[angle=0,height=4.1cm,width=0.95\textwidth,clip=]{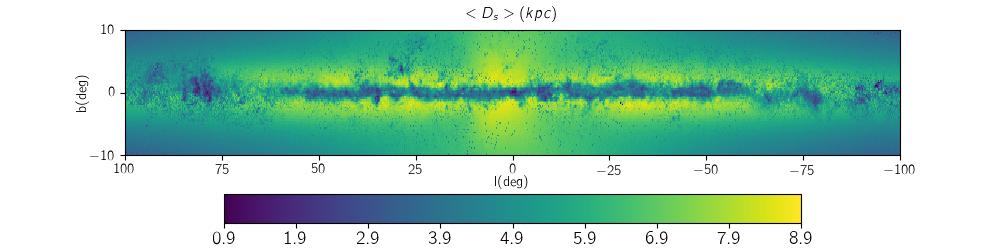}\label{fig2a}}
\subfigure[]{\includegraphics[angle=0,height=4.1cm,width=0.95\textwidth,clip=]{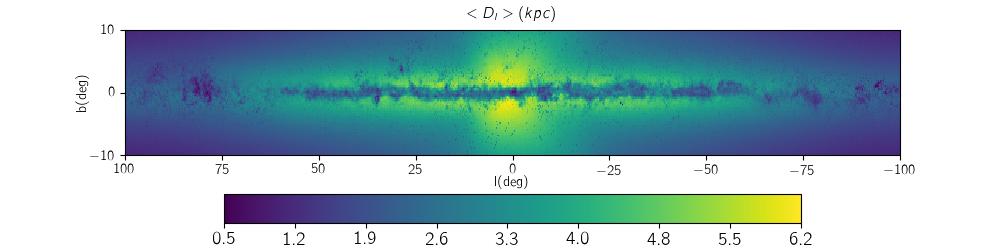}\label{fig2b}}
\subfigure[]{\includegraphics[angle=0,height=4.1cm,width=0.95\textwidth,clip=]{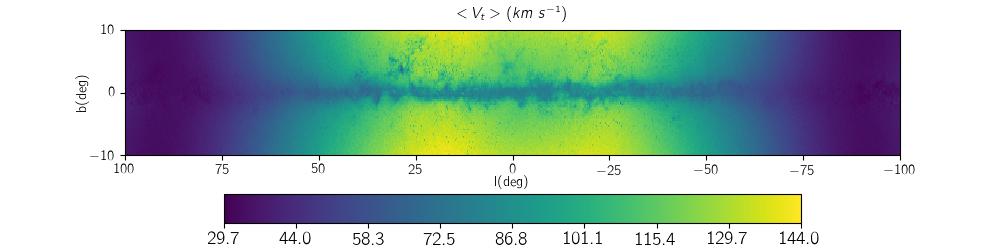}\label{fig2c}}
\subfigure[]{\includegraphics[angle=0,height=4.1cm,width=0.95\textwidth,clip=]{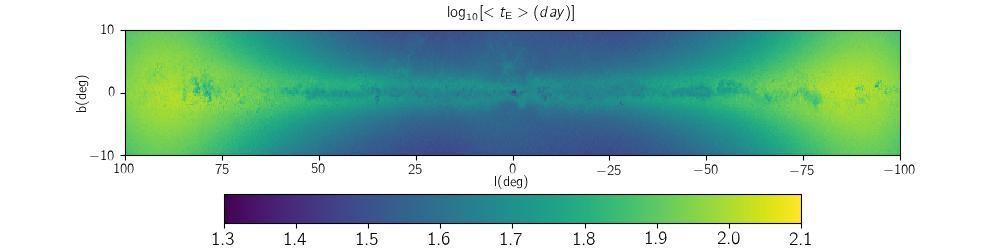}\label{fig2d}}
\caption{Maps of physical parameters for microlensing events
detectable by LSST toward the Galactic plane resulted from the Monte
Carlo simulation. The parameters are the source and lens distances
from the observer, the relative lens-source velocity and the
Einstein crossing time from top to bottom,
respectively.}\label{fig2}
\end{figure*}
Figures (\ref{fig1}) and (\ref{fig10}) show that the microlensing
events of very faint source stars (specially those fainter than the
LSST detection threshold) can not mostly be detected. These events
need to be highly magnified to be detected, whereas the durations of
high-magnification microlensing events are proportional to
$t_{\rm{E}}u_{0}$, mostly too short in comparison with the LSST
cadence. Whereas, the microlensing events of bright source stars
even with low magnification or long-duration ones are more likely to
be detected. For these events the time scale of the magnification is
of order of the Einstein crossing time, most likely long enough for
taking several data points by LSST. In the next section we study the
characteristics and statistical properties of the detectable
microlensing events.

\section{Observable Microlensing events with LSST}\label{four}
For each direction specified with the Galactic longitude and
latitude with the steps $\Delta(l)=\Delta(b)=0.25^{\rm{\circ}}$, we
perform a Monte Carlo simulation of microlensing events and probe
their detectability, so that for every direction we have an ensemble
of the detected events. In the following subsection we plot the map of
characterizations and statistics of these events to study them.
The numbers behind these maps are listed in Table (\ref{tab5}) and (\ref{tab6}) of the online version.

\begin{figure*}
\centering
\subfigure[]{\includegraphics[angle=0,height=4.1cm,width=0.95\textwidth,clip=]{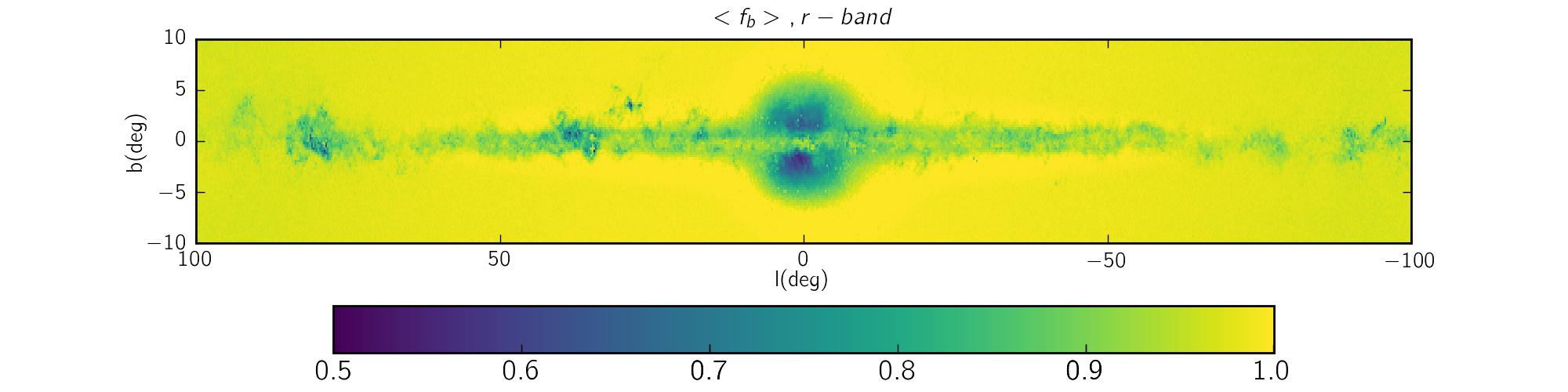}\label{fig2e}}
\subfigure[]{\includegraphics[angle=0,height=4.1cm,width=0.95\textwidth,clip=]{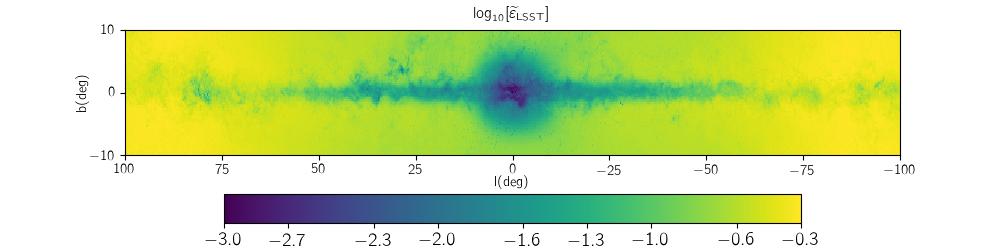}\label{fig3e}}
\caption{Maps of the blending parameter in $r$-band of observable
microlensing events (top) and the probability of detecting
microlensing events $\widetilde{\varepsilon}_{\rm{LSST}}(l,b)$ (bottom),
respectively.}\label{fig20}
\end{figure*}
\subsection{Characterizations and statistics} 
In Figure (\ref{fig2}), we plot the maps of the averaged physical
parameters of the lens and source stars for the observable
microlensing events. These parameters are the source and the lens
distances from the observer, the relative lens-source velocity, and
the Einstein crossing time from top to bottom.
Toward the small latitudes ($|b|<2^{\rm{\circ}}$), the interstellar
extinction is very high which causes mostly the microlensing events
with closer source and lens stars to be detectable, see Figure
\ref{fig2a} and \ref{fig2b}. According to Figure \ref{fig2b}, toward
the Galactic bulge excluding the points with $|b|<2^{\rm{\circ}}$
most of the lens stars belong to the Galactic bulge and are located
at the average distances around $6~\rm{kpc}$.

The Einstein crossing times of microlensing events far from the
Galactic bulge are intrinsically much longer than those toward the
Galactic bulge because the averaged relative lens-source velocity of
the events far from the Galactic bulge is much smaller than that of
the events toward the Galactic bulge, see Figures \ref{fig2c} and
\ref{fig2d}. For events that are far from the Galactic bulge on the
sky, the relative lens-source velocity is smaller because the lens
and source are close to each other and close to the Sun. Generally,
the detectable microlensing events with LSST (due to long cadence)
are a little longer than the detectable events with OGLE or MOA
surveys toward similar lines of sight. 

Figure (\ref{fig20}) represents two maps: the top one, shows the
blending parameter per line of sight averaged over the detectable
microlensing events in $r$-band. The blending parameter $f_{b}$ is a
function of the stellar number density and decreases while
increasing the stellar number density (plotted in Figure
\ref{fig0b}). In the bottom map, we plot the probability of
discerning microlensing events which their source stars are visible
at least at one filter toward a given direction,
$\widetilde{\varepsilon}_{\rm{LSST}}(l,b)$. For each line of sight,
this probability is the fraction of the simulated microlensing
events with detectable source stars (at least at one filter) which
pass the detectability criteria (mentioned in the subsection
\ref{ceriteria}). Generally, LSST detects the microlensing events
with longer durations than typical values. Far from the Galactic
bulge, the averaged Einstein timescale of microlensing events is
longer than that of the events toward the Galactic bulge. On the
other hand, the blending effect is ignorable toward the Galactic
disk. These two effects make the probability function
$\widetilde{\varepsilon}_{\rm{LSST}}(l,b)$ toward the Galactic plane
be more than that toward the Galactic bulge.

\begin{figure*}
\centering
\subfigure[]{\includegraphics[angle=0,height=4.05cm,width=0.95\textwidth,clip=]{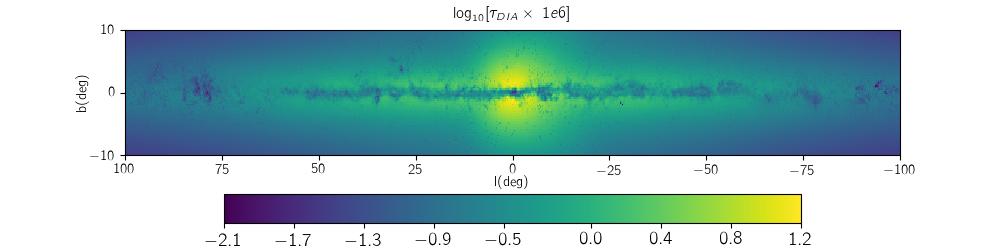}\label{fig3a}}
\subfigure[]{\includegraphics[angle=0,height=4.05cm,width=0.95\textwidth,clip=]{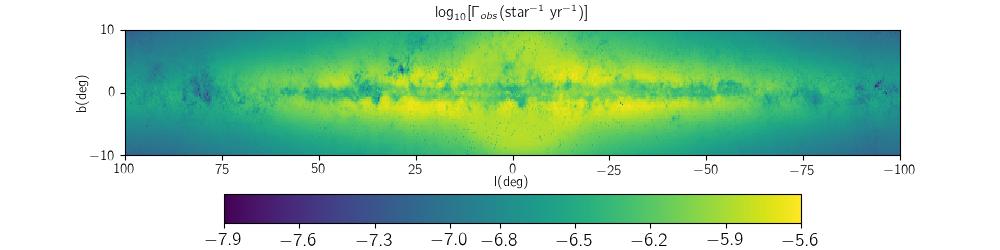}\label{fig3b}}
\subfigure[]{\includegraphics[angle=0,height=4.05cm,width=0.95\textwidth,clip=]{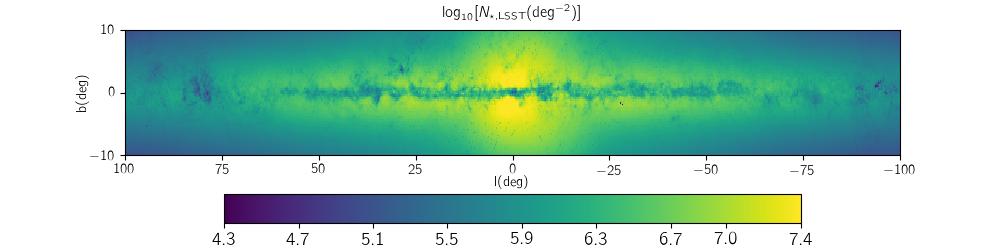}\label{fig3c}}
\subfigure[]{\includegraphics[angle=0,height=4.05cm,width=0.95\textwidth,clip=]{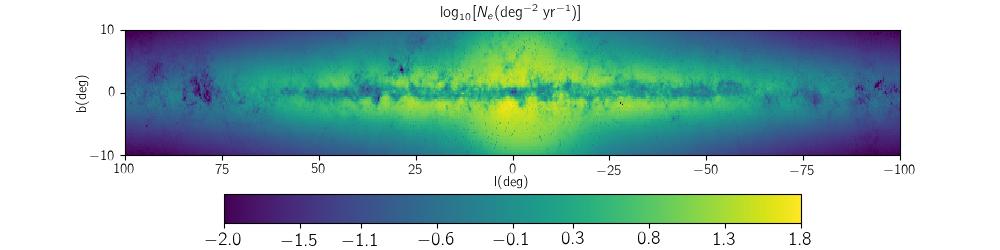}\label{fig3d}}
\caption{Statistics of microlensing parameters for LSST Galactic
plane events. These statistical parameters are the DIA optical
depth, the observing rate of events detectable by the LSST per year,
per star, the number of background stars detectable in at least one
of the LSST filters per square degree and the number of detected
microlensing events per year per square degree from top to bottom,
respectively. In the last panel, the values of $\log_{10}[N_{e}]$
smaller than $-2$ are marked same as $-2$ and the smallest recorded
value is $-3.5$. }\label{fig3}
\end{figure*}

For the statistics properties, we first calculate the overall
optical depth due to all structures for the simulated microlensing
events which are recognizable with LSST. The optical depth in given
direction and distance can be calculated as (e.g.,
\citet{Marc2017}):
\begin{eqnarray}\label{optical}
\tau_{\rm{DIA}}(l,b,D_{s})=\frac{4\pi~G~D^{2}_{s}}{c^{2}}\int^{1}_{0}~x(1-x)~\rho_{t}(l,b,x)~dx,
\end{eqnarray}
where $x=D_{l}/D_{s}$. Toward the Galactic disk, the distance of the
detectable sources has a wide distribution. In that case the optical
depth results from averaging over optical depths due to different
source distances \citep{Rahal2009}. We consider the source stars are
visible at peak (but not at baseline), the so-called DIA optical
depth \citep{Kerins2009}, $\tau_{\rm{DIA}}(l,b)$. The maps of the
DIA optical depth is shown in Figures \ref{fig3a}.
The DIA optical depth is reduced in $|b|<2^{\circ}$ region
due to the high extinction and blending, which prevent
detection of events with sources at large distances.

\begin{figure}
\begin{center}
\includegraphics[angle=0,width=8.5cm,clip=]{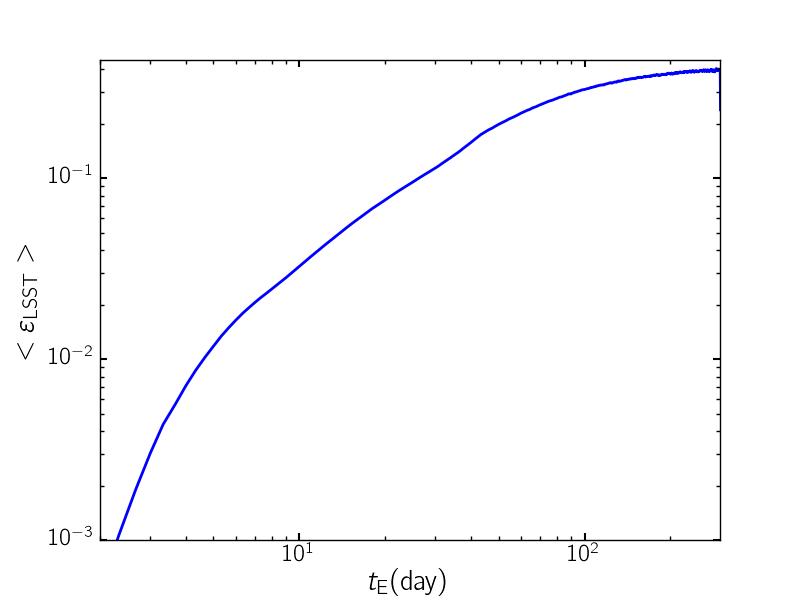}
\caption{The LSST efficiency for detecting microlensing event versus
the Einstein crossing time which is resulted from Monte Carlo
simulation.}\label{fig5}
\end{center}
\end{figure}

Using the DIA optical depths, we estimate the observed event rate
per line of sight as:
\begin{eqnarray}\label{rate}
\Gamma_{\rm{obs}}(l,b)=\frac{2}{\pi}\left<\frac{\varepsilon_{\rm{LSST}}(l,b,t_{\rm{E}})}{t_{\rm{E}}}\right>~\tau_{\rm{DIA}}(l,b),
\end{eqnarray}
where $\varepsilon_{\rm{LSST}}(l,b,t_{\rm{E}})$ is the LSST
efficiency for detecting microlensing events with the duration
$t_{\rm{E}}$ toward a given direction. Indeed, this function is the
mentioned probability function
$\widetilde{\varepsilon}_{\rm{LSST}}(l,b)$ for special group of
simulated microlensing events which have the timescale about
$t_{\rm{E}}$. The efficiency function $\varepsilon$ toward the
Galactic bulge and the Galactic disk with $|b|<2^{\rm{\circ}}$ is
much smaller than those toward the other directions, because of
their high blending effect and interstellar extinction, see Figure
\ref{fig3e}. In Figure (\ref{fig5}), the LSST efficiency averaged
over different directions $<\varepsilon_{\rm{LSST}}>$ versus the
Einstein crossing time is plotted. This efficiency is calculated via
a Monte-Carlo simulation over the whole Galactic plane. The
efficiency function increases with increasing the Einstein crossing
time. The event rate per line of sight through the Galaxy are
represented in Figure \ref{fig3b}. The event rate is high wherever
the stellar number density is high, except the directions with
$|b|<1.5^{\rm{\circ}}$, i.e., toward the Galactic disk with very
high interstellar extinction.

\begin{deluxetable*}{ccccccccccc}
\tablecolumns{11}\centering\tablewidth{0.9\textwidth}\tabletypesize\footnotesize\tablecaption{
Characteristics of detectable microlensing events for four different
lines of sights toward the Galactic plane.}\tablehead{
\colhead{$l,b$}&\colhead{$m_{base,r}$}&\colhead{$A_{r}$}&\colhead{$f_{b,r}$}&
\colhead{$\log_{10}[t_{\rm{E}}]$}&\colhead{$D_{s}$}&\colhead{$D_{l}$}&\colhead{$v_{t}$}&
\colhead{$\tau_{\rm{DIA}}(10^{-6})$}&\colhead{$N_{\star,l,\rm{LSST}}(10^{6})$}&\colhead{$N_{e,l}$}\\
$\rm{(deg,deg)}$& $\rm{(mag)}$ & $\rm{(mag)}$ & & $\rm{(day)}$ &
$\rm{(kpc)}$ & $\rm{(kpc)}$ & $\rm{(km~s^{-1})}$ & & & } \startdata
\\
$1.0,~-4.0$& $21.2$ & $1.3$  & $0.7$ & $1.43$ & $8.1$ & $5.6$ & $117.3$& $4.8$ &$1.5$& $19.0$ \\
                   &$1.6$  &$0.05$ &  $0.3$ & $0.32$ & $2.2$ & $1.9$ & $63.6$  &  &  & \\
\\
\hline\\
$26.5,~-2.25$&$21.5$ & $1.6$  &  $0.98$ & $1.52$ & $8.2$ & $4.6$ & $119.5$  & $1.1$ &$0.5$& $10.0$\\
             &$2.0$ &  $0.4$  &  $0.1$ & $0.33$ & $3.3$& $2.4$ & $64.2$  &  &  & \\
\\
\hline\\
$300.0,~3.0$& $21.8$ & $1.9$  &  $1.0$ & $1.69$ & $6.6$ & $3.0$ & $66.9$  & $0.3$ &$0.1$& $0.7$\\
                           & $2.2$ &  $0.4$  &  $0.0$ & $0.34$ & $3.2$& $1.8$ & $37.8$   & &  & \\
\\
\hline\\
$90.0,~5.0$& $21.9$ & $2.1$  &  $1.00$ & $1.87$ & $5.0$ & $1.9$ & $33.9$ & $0.09$ &$0.03$& $0.04$\\
                        &$2.3$ & $0.1$ &  $0.0$ & $0.30$ & $2.7$& $1.3$ & $20.6$ & &  & \\
\enddata
\tablecomments{For each direction, the averaged value of each
parameter and its standard deviation from the mean value are
mentioned in the first and second rows, respectively.
$N_{\star,l,\rm{LSST}}$ and $N_{e,l}$ are calculated over the area
$\Omega_{l}$.}\label{tab1}
\end{deluxetable*}

In order to estimate the number of observable microlensing events,
we evaluate the number of background and visible stars. Our
criterion for visibility is that blended stellar brightness should
be between LSST detection and saturation limits at least at one
filter. The total number of these stars toward a given direction and
per square degree is given by:
\begin{eqnarray}\label{nstar}
N_{\star,\rm{LSST}}(l,b)=\frac{2}{3}\int_{0}^{\infty}n_{t}(l,b,D_{s})\epsilon_{\rm{LSST}}(l,b,D_{s})D^{2}_{s}dD_{s},
\end{eqnarray}
where $\epsilon_{\rm{LSST}}(l,b,D_{s})$ is the efficiency for
specifying stars located at a given distance toward a given
direction at least at one LSST filter. We multiply
$N_{\star,\rm{LSST}}$ by a factor of $2/3$ because around one third
of stars are in binary systems and can not be discerned separately.
Figure \ref{fig3c} shows the
map of $N_{\star,\rm{LSST}}$ per square degree throughout the
Galactic plane, which is very high toward the Galactic bulge in
comparison with other directions. Because of the high extinction
toward the Galactic disk ($|b|<1.5^{\circ}$) the number of visible
stars reduces for these directions.

Finally, we can estimate the number of observable microlensing
events per square degree during the observing time $T_{obs}$, using:
\begin{eqnarray}\label{nevent}
N_{e}(l,b)=\Gamma_{\rm{obs}}(l,b)~T_{obs}~N_{\star,\rm{LSST}}(l,b),
\end{eqnarray}
Figure \ref{fig3d} represents the map of $N_{e}(l,b)$ per square
degree per year in the logarithmic scale. The number of microlensing
events toward the Galactic bulge and disk are in the order of $400$
and $15$ per square degree, respectively. Therefore, on the average
the LSST fields toward the Galactic bulge and disk the number of
events will be $3840$ and $144$, respectively during $10$ years
observation with the $3.9\rm{-day}$ cadence. Accordingly, LSST will
detect very large number of microlensing events in comparison with
the nowadays surveys because of (i) large FoV and (ii) its deep
observations. Certainly, the probability of observing special
events, e.g., the events with stellar black holes as microlenses, is
high in large population of detected microlensing events.
\begin{figure}
\subfigure[]{\includegraphics[angle=0,width=0.55\textwidth,clip=]{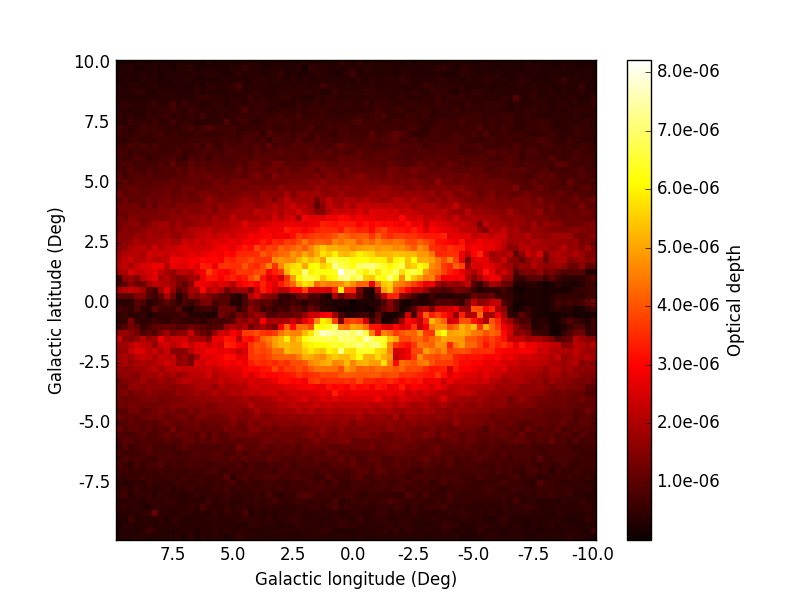}\label{app1}}
\subfigure[]{\includegraphics[angle=0,width=0.55\textwidth,clip=]{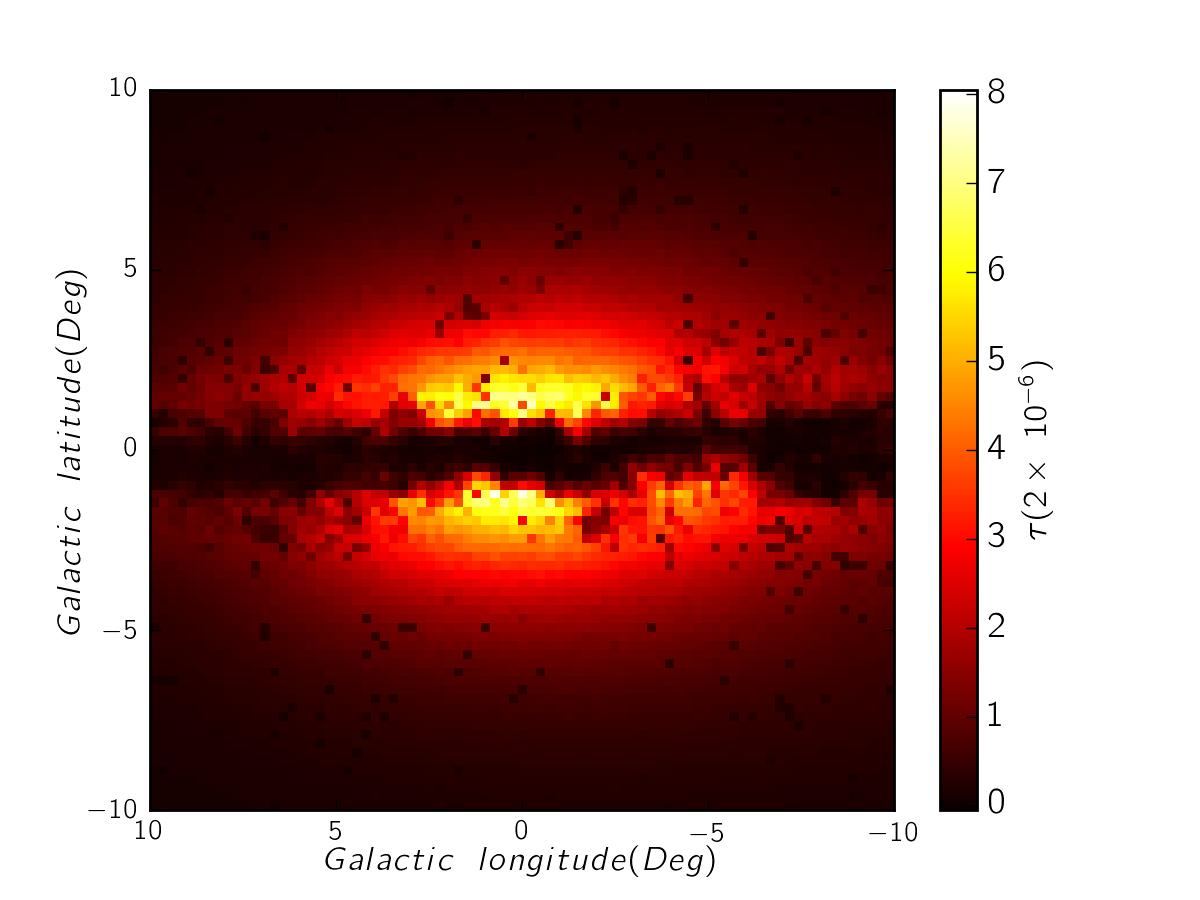}\label{app2}}
\caption{Top panel: The map of the optical depth for DIA sources
brighter than $23$ mag in $I$-band which was done by
Manchester-Besan\c{c}on Microlensing simulator(MaB$\mu$LS)
\citep{Awiphan2015}. Bottom panel: The map of optical depth resulted
from the simulation in this work. In order to have the same color
range for both plots, we multiply the bottom map by
$1/2$.}\label{append}
\end{figure}

In order to quantitatively compare the detectable microlensing
events toward different directions in the Galaxy, in Table
(\ref{tab1}), we report the characteristics and statistics of these
events for four different directions. The first column of this table
specifies the Galactic latitude and longitude of these directions.
The first direction is toward Baade's window, the second one is
toward the $\beta~$Sct ( which is one of four directions observed by
EROS-II, the end of the Galactic bar is visible from this line of
sight), other directions are toward the Galactic plane. In each row,
the values of the physical parameters (first rows) and their
standard deviations (second rows) calculated over the detectable
microlensing events are reported. Two last columns are calculated
over the area of the given line of sight, i.e.,
$N_{\star,l,\rm{LSST}}=N_{\star,\rm{LSST}}\times \Omega_{l}$ and
$N_{e,l}=N_{e}\times \Omega_{l}$. The number of visible stars per
line of sight and the blending effect toward the Galactic bulge is
higher than those toward the Galactic disk. As one can expect, the
optical depth toward the Galactic bulge is higher than that toward
the Galactic disk which causes the high number of detectable
microlensing events toward the Galactic bulge. In the next
subsection, we check the validity of our simulation by comparing
with other microlensing observations.

\subsection{Comparison of the LSST simulation with other observations} 
\begin{deluxetable*}{ccccccccc}
\tablecolumns{9}\centering
\tablewidth{0.85\textwidth}\tabletypesize\footnotesize\tablecaption{
The effect of changing cadence on the LSST microlensing detections
toward two directions with $(l=1^{\rm{\circ}},b=-4^{\rm{\circ}})$
and $(l=300^{\rm{\circ}},b=3^{\rm{\circ}})$.}
\tablehead{\colhead{$cadence$}&\colhead{$\log_{10}[t_{\rm{E}}]$}&
\colhead{$v_{t}$}&\colhead{$u_{0}$}&\colhead{$f_{b}$}&\colhead{$m_{base,r}$}&\colhead{$\widetilde{\varepsilon}_{\rm{LSST}}$}&\colhead{$\Gamma_{\rm{obs}}(10^{-7})$}&\colhead{$N_{e,l}$}\\
$\rm{(day)}$ & $\rm{(day)}$& $\rm{(km~s^{-1})}$& & & $\rm{(mag)}$& &
$\rm{(star^{-1}yr^{-1})}$& }\startdata
\\
&&&&$(l=1^{\rm{\circ}},b=-4^{\rm{\circ}})$&&&\\
\hline\hline\\

$2.0$& $1.37$ & $127.10$ & $0.42$ & $0.68$ & $21.50$ & $2.93e-3$ & $21.59$ &$31.18$ \\
\hline\\
$3.0$& $1.40$ & $122.13$ & $0.43$ &  $0.71$ & $21.33$ & $2.29e-3$ & $15.92$ & $23.33$\\
\hline\\
$4.0$& $1.43$ & $117.21$ & $0.44$ &  $0.73$ & $21.21$ & $1.83e-3$ & $12.51$& $18.26$ \\
\hline\\
$5.0$ & $1.46$ & $112.81$ & $0.44$ &  $0.74$ & $21.11$ & $1.49e-3$ & $10.19$ & $14.88$\\
\hline\\
$6.0$ & $1.49$ & $108.39$ & $0.45$ &  $0.76$ & $21.03$ & $1.24e-3$ & $8.52$ & $12.17$\\
\hline\\
&&& &$(l=300^{\circ},b=3^{\circ})$&&&\\
\hline\hline\\
$2.0$& $1.65$ & $70.40$ & $0.38$ & $1.00$ & $22.07$ & $0.42$ & $7.81$ &$1.00$ \\
\hline\\
$3.0$& $1.68$ & $68.60$ & $0.44$ &  $1.00$ & $21.89$ & $0.36$ & $6.42$ & $0.82$\\
\hline\\
$4.0$& $1.70$ & $66.54$ & $0.44$ &  $1.00$ & $21.75$ & $0.32$ & $5.45$& $0.70$ \\
\hline\\
$5.0$ & $1.74$ & $65.01$ & $0.45$ &  $1.00$ & $21.64$ & $0.29$ & $4.73$ & $0.60$\\
\hline\\
$6.0$ & $72.46$ & $63.46$ & $0.45$ &  $1.00$ & $21.55$ & $0.26$ &
$4.18$ & $0.53$
\enddata
\label{tab2}
\end{deluxetable*}

The EROS-II team has searched the microlensing events toward the
Galactic spiral arms, away from the Galactic bulge
\citep{Rahal2009}. \citet{Marc2017} simulated the EROS-II
observations and from comparing the real observations with the
simulation constrained the kinematics of the disk, the stellar mass
function and the maximum contribution of a thick disk in the form of
the compact objects. In order to test our code by replicating the
results of EROS-II, we repeat the simulation toward the four
directions of EROS-II observations with the same conditions. Also,
we use the function of the EROS-II efficiency for detecting
microlensing events versus the Einstein crossing time (plotted in
their Figure (6)) as the detectability function. Toward $\beta~$Sct
with the coordinates $l=26.6^{\circ},b=-2.2^{\circ}$ we get the
average value of the Einstein crossing time of $56.9$ days, which is in
the agreement with the observed value of $59 \pm 31$ days. The
quoted uncertainty is the standard deviation from the mean value.
Toward $\gamma~$Sct, $l=18.5^{\circ},b=-2.1^{\circ}$, we get
$<t_{\rm{E}}>=52 \pm 30$ days, while the observed value was $47\pm
32$ days. Toward $\gamma~$Nor with $l=331.1^{\circ},b=-2.4^{\circ}$,
from simulation $<t_{\rm{E}}>=54$ days, again in the same order of
the observation amount, i.e. $57$ days. Finally, toward $\theta~$Mus
with $l=306.6^{\circ},b=-1.5^{\circ}$ $<t_{\rm{E}}>$ from simulation
and observations are $81\pm 49$ and $97$ days. The small differences
between the results from our simulations and real observation are
due to difference photometry systems for the source stars, their
observing strategies, etc.

We also compare the optical depth map from this simulation with one
resulted from Manchester-Besan\c{c}on Microlensing Simulator
(MaB$\mu$lS)\footnote{\url{http://www.mabuls.net/}} which was developed by
\citet{Awiphan2015}. This simulation was based on the MOA-II
observations of microlensing events. The map of the optical depth
for DIA sources brighter than $23$ mag in $I$-band (the faintest
limit in their simulation) from MaB$\mu$lS is shown in Figure
\ref{app1} and our optical depth map in the same range of the
Galactic latitude and longitude is plotted in Figure \ref{app2}.
These two maps are similar, but the optical depth from LSST
simulation is larger than that from MaB$\mu$LS. One reason is that
LSST is deeper than $23$ mag in $I$-band, even by considering the
difference in their exposure times (the exposure time of MOA-II
observations is $60$ seconds twice the LSST exposure which results
the difference in their limiting magnitude diminishes by $\sim0.3$
mag).

\subsection{The impact of the LSST cadence on the Microlensing observation}
In order to study the effect of improving the cadences on the
microlensing detections with LSST, we perform the simulation with
several cadences equal to $(2.0,3.0,4.0,5.0,6.0)$ days. In table
(\ref{tab2}), we report the physical parameters that depend on the
cadence. These simulations are done for two directions toward (i)
the Baade's window with $(l=1^{\rm{\circ}},b=-4^{\rm{\circ}})$ and
(ii) the Galactic disk with $(l=300^{\rm{\circ}},b=3^{\rm{\circ}})$.
According to the table, improving the cadence makes the
shorter-duration microlensing events due to somewhat fainter source
stars be more observed. Since, the number of these events are
intrinsically high, so by improving the cadence the number of
detectable microlensing events raises. According to Table
(\ref{tab2}), changing the cadence from $6$ days to $2$ days doubles
the number of observable microlensing events throughout the Galaxy.


\section{LSST observing strategies}\label{five}
\begin{figure*}
\begin{center}\includegraphics[angle=0,width=0.9\textwidth,clip=]{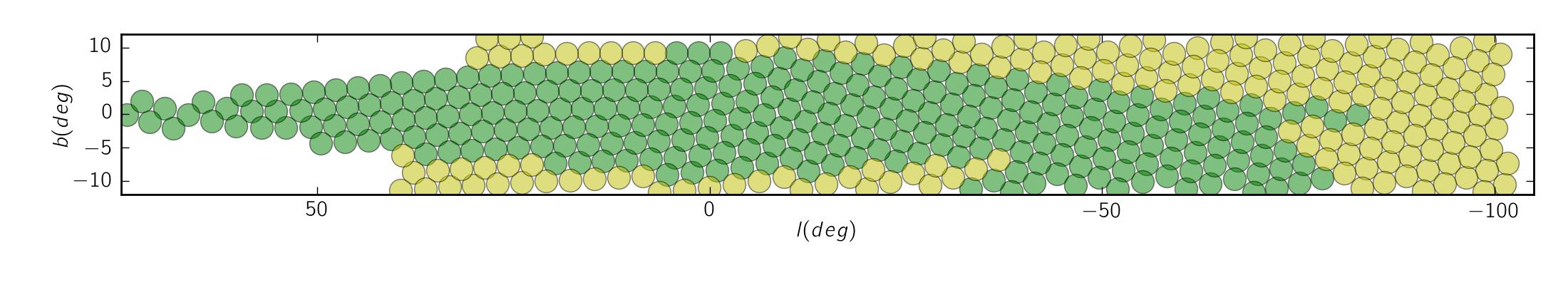}
\caption{The fields will be detected by LSST (I) during $10$ years
with around $900$ epochs and with the averaged $3.9\rm{-day}$
cadence (yellow circles) and (II) during the first year with $30$
epochs in each filter (green circles). The area of each field is
$9.6~ \rm{deg^{2}}$.}\label{field}
\end{center}
\end{figure*}

LSST is supposed to observe $449$ fields of the Galactic plane,
i.e., with the coordinates $|b|<10^{\rm{\circ}}$ and
$|l|<100^{\rm{\circ}}$. Each field is almost circle-shape with the
diameter $3.5~\rm{deg}$ and the area $9.6~\rm{deg^{2}}$ which
corresponds to $150$ lines of sight in our simulations. $178$ of
these fields will be observed by LSST with the $3.9\rm{-day}$
cadence during $10$ years (I) and $271$ of these fields LSST will be
observed only during the first year and with a $0.96\rm{-day}$
cadence (II). These fields are shown in Figure (\ref{field}). The
fields will be observed with the $3.9\rm{-day}$ cadence are
represented with yellow circles and the others are shown with green
ones.

\begin{deluxetable*}{cccccccccccc}
\tablecolumns{12}\centering\tablewidth{0.9\textwidth}\tabletypesize\footnotesize\tablecaption{
Characteristics of detectable microlensing events toward the
Galactic plane with LSST by considering two different strategies (I)
and (II), resulted from the Monte Carlo simulation.}
\tablehead{\colhead{$~~$}&\colhead{$No.~Epochs$}&
\colhead{$No.~Fields$}&\colhead{$cadence$}&\colhead{$T_{obs}$}&\colhead{$<\log_{10}[t_{\rm{E}}]>$}&
\colhead{$<u_{0}>$}&\colhead{$<V_{t}>$}&\colhead{$<f_{b,r}>$}&\colhead{$<A_{r}>$}&\colhead{$<\Gamma_{\rm{obs}}(10^{-7})>$}&\colhead{$N_{e,sum}$}\\
 & & & $\rm{(day)}$ & $\rm{(yr)}$ & $\rm{(day)}$ & & $\rm{(km~s^{-1})}$ &  & $\rm{(mag)}$ & $\rm{(star^{-1}~yr^{-1})}$& }
\startdata
\\
$(I)$ & $900$ & $178$ & $3.92$ & $10.0$ &  $1.81$ & $0.44$ & $79.10$ & $0.94$ & $1.19$& $4.28$ & $7954.9$ \\
\hline\\
\\
$(II)$ & $180$ & $271$ & $0.96$ & $1.0$& $1.63$ & $0.42$ & $110.21$
& $0.82$ & $3.08$ & $23.68$ & $34155.3$
\enddata
\tablecomments{Averaging is done over the detectable microlensing
events toward different fields. The last column is the summation
number of detectable events over all related fields.}\label{tab3}
\end{deluxetable*}
\begin{deluxetable*}{ccccccc}
\tablecolumns{7}\centering\tablewidth{0.9\textwidth}\tabletypesize\footnotesize\tablecaption{The
impact of two LSST observing strategies, i.e., (I) and (II), on
characteristics and statistics of detectable microlensing events
toward $390$ different fields. The first column, $(l_{c},b_{c})$,
indicates the coordinates of the center of each field. For other
columns and for each row, two values are given which are
corresponding to the strategies (I) and (II), respectively. The
averaging is done over the area of each field.}
\tablehead{\colhead{$l_{c},b_{c}$}&\colhead{$<\log_{10}(t_{\rm{E}})>$}&
\colhead{$<m_{base,r}>$}&\colhead{$<f_{b}>$}&\colhead{$\log_{10}[\Gamma_{\rm{obs}}(10^{7})]$}&\colhead{$N_{e}$}&\colhead{$\log_{10}[<\widetilde{\varepsilon}_{\rm{LSST}}>]$}\\
$\rm{(deg,deg)}$ & $\rm{(day)}$ & $\rm{(mag)}$&&$\rm{(star^{-1}
yr^{-1})}$& $\rm{(deg^{-2})}$ &} \startdata
$-99.02,-8.39$ & $1.75,1.70$ & $20.32,19.94$ & $0.93,0.90$ & $0.12,0.33$ & $0.37,0.06$ & $-0.20,-0.08$ \\
$-99.53,-5.24$ & $1.79,1.74$ & $20.01,19.31$ & $0.90,0.86$ & $0.29,0.50$ & $1.08,0.17$ & $-0.22,-0.10$ \\
$-99.81,+6.04$ & $1.78,1.73$  & $20.35,19.90$ & $0.92,0.89$ & $0.26,0.46$ & $0.37,0.06$ & $-0.21,-0.09$ \\
$-98.14,-0.07$ & $1.83,1.78$  & $16.68,15.03$ & $0.72,0.65$ & $0.44,0.66$ & $0.94,0.16$ & $-0.29,-0.15$ \\
$-98.94,+3.00$ & $1.82,1.77$ &  $19.79,18.90$ & $0.88,0.83$ & $0.45,0.64$ & $0.95,0.15$ & $-0.23,-0.11$ \\
$-97.44,-3.18$ & $1.83,1.78$ &  $19.41,18.47$ & $0.86,0.81$ & $0.40,0.61$ & $1.95,0.32$ & $-0.24,-0.11$ \\
$-97.79,+8.01$ & $1.77,1.71$ &  $20.27,19.94$ & $0.92,0.90$ & $0.14,0.35$ & $0.19,0.03$ & $-0.20,-0.08$ \\
$-96.35,-9.44$ & $1.76,1.71$ &  $20.32,19.98$ & $0.93,0.90$ & $0.08,0.29$ & $0.28,0.05$ & $-0.19,-0.08$ \\
$-96.85,-6.31$ & $1.79,1.74$ &  $20.20,19.64$ & $0.92,0.88$ & $0.24,0.44$ & $0.79,0.13$ & $-0.21,-0.09$ \\
$-96.10,+1.95$ & $1.85,1.80$ & $18.00,16.65$ & $0.79,0.72$ & $0.44,0.65$ & $0.77,0.13$ & $-0.26,-0.13$ \\
$-96.91,+4.99$ & $1.82,1.76$ & $20.24,19.75$ & $0.92,0.88$ & $0.33,0.53$ & $0.60,0.10$ & $-0.20,-0.09$ \\
$-95.36,-1.13$ & $1.85,1.80$ & $18.42,17.21$ & $0.81,0.75$ & $0.49,0.70$ & $1.72,0.30$ & $-0.25,-0.12$ \\
$-95.72,+9.97$ & $1.76,1.71$ & $20.19,19.80$ & $0.92,0.89$ & $0.06,0.27$ & $0.16,0.03$ & $-0.19,-0.08$ \\
$-94.15,-7.34$ & $1.80,1.75$ & $19.92,19.33$ & $0.90,0.86$ & $0.16,0.36$ & $0.52,0.08$ & $-0.21,-0.09$ \\
$-94.71,-4.23$ & $1.84,1.79$ & $20.16,19.55$ & $0.91,0.87$ & $0.38,0.58$ & $2.04,0.32$ & $-0.20,-0.09$ \\
\enddata
\tablecomments{A complete electronic version of this table is
available at: ....}\label{tab4}
\end{deluxetable*}

We used LSST OpSim simulations to extract epoch of observation,
airmass, seeing FWHM, and filter for each visit in each field
separately. These parameters have impact on the results of the
simulation. We perform the Monte Carlo simulation for all of these
fields to estimate the number of microlensing events that are
detected by LSST by considering the corresponding their realistic
strategies (I) and (II) and use the sequence of time, airmass, FWHM,
filter for simulating syntectic data points for each simulated light
curve. The table (\ref{tab3}) contains the results of these
simulations. The first row represents the characteristics of
detectable microlensing events toward the fields shown with yellow
circle in Figure (\ref{field}) during $10$ years observations with
the $3.9\rm{-day}$ cadence and the second row, (II), shows the
properties of the detectable microlensing events toward the other
fields (green circle in Figure \ref{field}).

According to this table, LSST with the second strategy (II) will
detect shorter-duration events than those are detectable with the
first strategy (I). According to Figure (\ref{field}), most of
fields with small Galactic latitude (toward the Galactic bulge and
disk) will be detected with the second strategy, whereas toward
these directions the optical depth is higher. Consequently, the
number of detectable microlensing events with the strategy (II)
during one year with the $0.96\rm{-day}$ cadence is more than those
detected by the first strategy. Also, the stellar extinction for
these directions is high which makes the averaged impact parameter
of the detectable events be a little smaller. However, the
microlensing events potentially detectable the strategy (II)
will be of low value.
The reasons are (i) the number of baseline data points is
rare so that the source fluxes can not be estimated accurately. (ii)
Lack of baseline data intensifies the degeneracy between the source
flux and the Einstein crossing time which in turn causes weak
estimations of the events' timescale. During one year observation,
(iii) measuring the parallax effect for long-duration microlensing
events and (iv) discerning the variable stars from microlensing
events are difficult. All of these issues can be solved by
increasing the observational time to longer than one year.

Here, we compare these strategies in the regard of detecting
microlensing events toward similar fields, by simulating detectable
microlensing events. Table (\ref{tab4}) contains some characteristic
and statistic parameters of detectable microlensing events toward
different fields by considering the LSST observing strategies (I)
and (II) (the first and second values of each parameter are due to
strategies I and II, respectively). This table can be used to
optimize LSST observing strategy for disk microlensing events. As an
example, when some of the fields currently scheduled for strategy
(II) could be observed with strategy (I), one can select the bins of
Galactic longitude and in each bin select a field for which the
change of strategy would give the highest increase in $N_{e}$. We
note that the ratio of $N_{e}$ values for any given field in the two
strategies is less dependent on simulation details than the raw
values themselves. According to this table:

\begin{itemize}
\item{Generally, during $10$ years observation with $900$ epochs (I)
LSST will detect more and on average longer microlensing events than
those detectable during one year observation with $180$ epochs (II).
Detecting longer microlensing events somewhat justifies the negative
effect of long cadence in the observing strategy (I) on detection of
short signals in microlenssing events, e.g., planetary ones.
Usually, the longer microlensing events have longer planetary
signals, because the time of caustic crossing is proportional to the
Einstein crossing time. Hence, with the first observing strategy (I)
although the time interval between data points is long and around
$3.9$ days, but detectable events are on average longer and as a
result have longer planetary signals.}

\item{With the second strategy (II) the probability of detecting
microlensing events is higher than that with the other, because the
number of short-duration microlensing events is intrinsically higher
than long-duration ones. This cause that increasing the observing
time from one year to $10$ years does not enhance the number of
detectable events by a factor of $10$.}

\item{Mostly, the source stars of detectable events with the strategy
(I) are somewhat fainter with higher blending effect than those with
the second strategy.}

\end{itemize}

Accordingly, LSST can have significant sensitivity to exoplanets
throughout the Galaxy, if either if cadence better than $\sim1-$day
is executed, or follow-up observations are conducted.
By detecting planets toward different
directions in the Galaxy, we can study their Galactic distribution \citep{Gould2013}.

\section{Summary and conclusions}\label{six}
We studied the detection of microlensing events toward the Galactic
disk and bulge with LSST. In this regard, we performed a Monte Carlo
simulation according to its strategy toward the Galactic longitude
and latitude in the ranges of $|l|<100^\mathrm{\circ}$ and
$|b|<10^\mathrm{\circ}$. We assumed that the cadence of the LSST
observations is equal to $3.9$ days and its exposure time is $30$
seconds. The results of the simulation were

(i) LSST mostly detects the microlensing events of source stars with
the average magnitude around $22$ in $r-$band. Although fainter
stars (up to $24.3$ mag in this filter) are visible by LSST, but
microlensing events due to these faint stars have small chance to be
realized. Because of large blending effect they have to be highly
magnified to generate high enough signal to noise ratios. But on the
other hand, the high-magnification microlensing events' durations
are scaled by the lens impact parameters, i.e., their durations are
mostly short in comparison to the LSST cadence.

(ii) Generally, the detectable microlensing events are (little)
longer than common events observable with nowadays surveys. Thus,
LSST partly helps to study microlensing events with more massive
microlenses or closer events, etc.

(iii) We predicted that LSST on average detects around $400$ and
$15$ microlensing events per square degree (or $3840$ and $144$ per
LSST's FoV) during its lifetime toward the Galactic bulge and disk,
respectively. This large number of visible microlensing events are
due to its large FoV and its observing depth.

We performed some simulations with different cadences to study the
effect of cadence on the statistics and properties of observable
microlensing events. Our simulations show that improving the cadence
increases the detection efficiency for short-timescale events. The
number of these events is intrinsically high. Therefore improving
the cadence raises the number of detectable microlensing events,
e.g. improving the cadence from $6$ days to $2$ days approximately
doubles the number of detectable microlensing events throughout the
Galaxy.

The current strategy for LSST is that it observes (I) some parts of
the sky with the average $3.9\rm{-day}$ cadence during $10$ years
and (II) some other parts of the sky with $180$ epochs during the
first year of its lifetime. We performed the simulation by
considering the corresponding strategies for these fields (shown in
Figure \ref{field}) and concluded that the number of events
corresponding to these strategies are $7900$ and $34000$,
respectively. Most of fields with small Galactic latitude (toward
the Galactic bulge and disk) are supposed to be detected with the
second strategy. Toward these directions, the stellar number density
and as a results the optical depth and the event rate are higher
which results the larger number of detectable microlensing events.

We have also presented expected number of events for each field
under both observing strategies (Table \ref{tab4}) which could be
used to optimize the LSST observing strategy. Toward alike field,
LSST with first observing strategy (I) will detect more and on
average longer microlensing events than those observable with the
strategy (II). Although the cadence in the first strategy is long,
but on the other hand long observing time ($10$ years) helps
carefully measuring the baseline source fluxes and the parallax
effect for long events, discerning the variable stars from
microlensing events, etc.,  whereas they are difficult to measure if
only one year of observations is available. In addition, because of
longer observing time the statistic of detectable events is higher
with the first strategy than that with the second strategy.

Lastly, LSST can have significant role on detecting planets
throughout the Galaxy and studying their Galactic distribution;
either with the first strategy as well as follow-up observations or
with the second strategy, i.e. $\sim 1-$day cadence or shorter, but
with somewhat longer observing time. In addition, LSST Observation
of the Galactic disk during several years with $\sim3.9-$day
cadence, first strategy (I), will allow finding isolated black
holes. Long observing time helps to measure their parallax effects
and baseline magnitudes, etc.

\begin{acknowledgments}
We thank Mike Lund for consultation. We especially acknowledges M.
Penney for careful reading and commenting on the manuscript. The
work by S. Sajadian was supported by a grant (95843339) from the
Iran National Science Foundation (INSF).
\end{acknowledgments}
\bibliographystyle{apj}
\bibliography{references}

\appendix
We present here Table (\ref{tab6}) and (\ref{tab5}) which contain the number behind
the maps shown in Figure (\ref{fig0}), (\ref{fig2}), (\ref{fig20}) and (\ref{fig3}).
\begin{deluxetable*}{ccccccc}
\tablecolumns{7}\centering
\tablewidth{0.6\textwidth}\tabletypesize\footnotesize\tablecaption{Properties of detectable microlensing events with LSST, resulted from Monte Carlo simulation in this work. 
}
\tablehead{\colhead{$b$}&\colhead{$l$}&\colhead{$filter$}&\colhead{$m_{base}$}&\colhead{$A$}&\colhead{$f_{b}$}&\colhead{$\log_{10}[N_{\rm{PSF}}]$}\\
$\rm{(deg)}$ & $\rm{(deg)}$ &  & $\rm{(mag)}$ & $\rm{(mag)}$ & & }
\startdata
$-10.00$ &  $-100.00$ &   $u$  &  $21.69970$  &  $0.64350$  &  $0.39450$ &   $0.00000$ \\
$-10.00$ &  $-100.00$ &   $g$  &  $22.81030$  &  $0.48670$  &  $0.90100$ &   $0.00000$ \\
$-10.00$ &  $-100.00$  &  $r$  &  $22.34540$  &  $0.35430$  &  $0.97950$ &   $0.00000$ \\
$-10.00$ &  $-100.00$ &   $i$  &  $21.72710$  &  $0.26670$  &  $0.99450$ &   $0.00000$ \\
$-10.00$ &  $-100.00$ &   $z$  &  $21.15100$  &  $0.19060$  &  $0.99800$ &   $0.00000$ \\
$-10.00$ &  $-99.75$  &  $u$   & $21.43780$   & $1.12170$ &  $0.40550$   &   $0.00000$ \\
$-10.00$ &  $-99.75$  &  $g$   & $22.71040$   & $0.84760$ &   $0.87600$  &  $0.00000$\\
$-10.00$ &  $-99.75$  &  $r$   &  $22.31460$   & $0.61890$ &  $0.96700$  &  $0.00000$\\
$-10.00$ &  $-99.75$  &  $i$   & $21.67870$   & $0.46510$  &  $0.99050$  &  $0.00000$ \\
$-10.00$ &  $-99.75$  &  $z$   &  $21.11810$  &  $0.33260$  &  $0.99200$ &   $0.00000$ \\
$-10.00$ &  $-99.50$ &  $u$ &  $21.80940$ &  $1.39950$ &  $0.37000$ &  $0.00000$ \\
$-10.00$ &  $-99.50$ &  $g$ &  $22.90770$ &  $1.05950$ &  $0.85450$ &  $0.00000$ \\
$-10.00$ &  $-99.50$ &  $r$ &  $22.50760$ &  $0.77520$ &  $0.97600$ &  $0.00000$ \\
$-10.00$ &  $-99.50$ &  $i$ &  $21.85850$ &  $0.58320$ &  $0.99400$ &  $0.00000$ \\
$-10.00$ &  $-99.50$ &  $z$ &  $21.24650$ &  $0.41770$ &  $0.99650$ &  $0.00000$\\
\enddata
\tablecomments{A complete electronic version of this table is available at: ....}
\label{tab6}
\end{deluxetable*}

\clearpage

\begin{turnpage}

\thispagestyle{empty} 

\begin{deluxetable*}{cccccccccccccc}[tb]
\tablecolumns{14}\centering
\tablewidth{0.95\textheight}\tabletypesize\footnotesize\tablecaption{Characteristics and statistics of detectable microlensing events with LSST by considering $3.9\rm{-day}$ cadence during $10$ years observation, resulted from the Monte Carlo simulation in this work.}
\tablehead{\colhead{$b$}&\colhead{$l$}&\colhead{$\log_{10}[\widetilde{\varepsilon}_{\rm{LSST}}]$}&\colhead{$t_{\rm{E}}$}&\colhead{$R_{\rm{E}}$}&\colhead{$D_{s}$}&\colhead{$D_{l}$}&\colhead{$V_{t}$}&\colhead{$u_{0}$}&\colhead{$\log_{10}[M_{t,l}]$}&\colhead{$\log_{10}[\Gamma_{\rm{obs}}(10^{7})]$}&\colhead{$\log_{10}[N_{e}]$}&\colhead{$\log_{10}[N_{\star,LSST}]$}&\colhead{$\log_{10}[\tau_{\rm{DIA}}(10^{6})]$}\\
$\rm{(deg)}$ & $\rm{(deg)}$ &  & $\rm{(day)}$ & $\rm{(A.U.)}$ & $\rm{(kpc)}$ & $\rm{(kpc)}$ & $\rm{(km~s^{-1})}$ &  & $\rm{(M_{\odot})}$ & $\rm{(star^{-1}~yr^{-1})}$ & $\rm{(deg^{-2}~yr^{-1})}$ & $\rm{(deg^{-2})}$ &}
\startdata
$-10.00$ & $-100.00$ &  $-0.32950$ & $74.01988$ &  $1.22901$ & $3.46638$ & $1.15046$ & $37.13148$ & $0.44591$ & $3.90800$ & $-0.09399$ & $-1.97754$ &  $5.11645$ & $-1.46942$\\
$-10.00$ & $-99.75$ & $-0.34193$ & $74.53259$ & $1.23582$ & $3.47254$ & $1.17777$ & $37.51582$ & $0.45630$ & $3.90986$ &
$-0.11888$ & $-2.00431$ & $5.11457$ & $-1.46623$\\
$-10.00$ & $-99.50$ & $-0.35131$ & $72.55770$ & $1.18267$ & $3.37450$ &  $1.13198$ & $36.78578$ & $0.43920$ & $3.91171$ & $-0.11623$ & $-2.01914$ & $5.09709$ & $-1.47847$\\
$-10.00$ & $-99.25$ & $-0.35372$ & $75.76251$ & $1.21863$ & $3.48016$ & $1.13086$ & $36.20193$ & $0.43940$ & $3.91358$ &
$-0.12951$ & $-2.02424$ & $5.10527$ & $-1.46433$\\
$-10.00$ & $-99.00$ & $-0.31723$ & $78.02045$ & $1.26823$ & $3.59088$ & $1.19090$ & $36.53486$ & $0.45678$ & $3.91545$ &
$-0.09689$ & $-1.96167$ &  $5.13522$ &  $-1.44782$ \\
$-10.00$ & $-98.75$ & $-0.30567$ & $76.64276$ & $1.24093$ & $3.55187$ & $1.15963$ & $35.66640$ & $0.45350$ & $3.91732$ &  $-0.08422$ & $-1.94408$ & $5.14013$ & $-1.45223$ \\
$-10.00$ & $-98.50$ & $-0.32552$ & $75.24845$ & $1.23505$ & $3.47227$ & $1.15603$ &  $36.39649$ & $0.45534$ & $3.91919$ & $-0.10268$ & $-1.98341$ & $5.11927$ & $-1.46294$ \\
$-10.00$ & $-98.25$ & $-0.30856$ & $75.94121$ & $1.23738$ & $3.50554$ & $1.17161$ & $36.38862$ & $0.46233$ & $3.92107$  & $-0.08374$ & $-1.93576$ & $5.14798$ & $-1.45383$ \\
$-10.00$ &  $-98.00$ & $-0.31838$ & $77.03021$ & $1.24994$ & $3.54215$ & $1.16636$ & $36.09435$ & $0.44784$ & $3.92296$ &
$-0.09162$ & $-1.94100$ & $5.15062$ & $-1.44843$ \\
$-10.00$ & $-97.75$ & $-0.31122$ & $78.53271$ & $1.25469$ & $3.54835$ & $1.17180$ & $35.98262$  & $0.44922$ & $3.92485$ & $-0.07587$ & $-1.93779$ & $5.13808$ & $-1.44928$ \\
$-10.00$ & $-97.50$ & $-0.30685$ & $75.28133$ & $1.22850$ & $3.42993$ &$1.15033$ & $36.34997$ & $0.45606$& $3.92674$ &    $-0.07189$ & $-1.92490$ & $5.14699$&  $-1.46231$ \\
$-10.00$  &$-97.25$ & $-0.33082$ & $79.04265$ & $1.20851$ & $3.50570$& $1.16402$ & $34.02276$ & $0.45836$& $3.92864$ &    $-0.12090$&  $-1.98132$ & $5.13958$ & $-1.45210$ \\
$-10.00$  &$-97.00$  & $-0.31608$ & $81.40951$ & $1.26586$ & $3.60313$&  $1.19656$ & $34.78856$ & $0.45368$& $3.93054$&    $-0.11019$  &$-1.97297$ & $5.13722$ & $-1.44009$ \\
$-10.00$ & $-96.75$ & $-0.29546$ & $76.32363$ & $1.22607$ & $3.56775$ & $1.17462$ & $35.93456$ & $0.45508$ & $3.93245$ &   $-0.06069$ & $-1.90376$ & $5.15693$ & $-1.44033$\\
$-10.00$ & $-96.50$ & $-0.31639$ & $79.89724$ & $1.24175$ & $3.54047$ & $1.16884$ &  $34.20550$ &  $0.45662$ &  $3.93436$ &  $-0.11107$ &  $-1.95927$ &  $5.15180$ &  $-1.44527$ \\
$-10.00$ & $-96.25$ & $-0.31503$ & $79.64673$ & $1.23656$ & $3.52473$ & $1.18407$ & $34.88030$ &  $0.45795$ &  $3.93628$ &  $-0.09068$ &  $-1.92447$ &  $5.16621$ &  $-1.44406$ \\
 $-10.00$ & $-96.00$ & $-0.29148$ & $80.02194$ & $1.25062$ & $3.57555$ & $1.20952$  & $34.98413$ &  $0.45785$ &  $3.93820$ &  $-0.07200$ &  $-1.91518$ &  $5.15682$ &  $-1.43938$ \\
$-10.00$ & $-95.75$ & $-0.29907$ & $78.89009$ & $1.23438$ & $3.54191$ & $1.17186$ &  $34.19057$ &  $0.44439$ &  $3.94012$ &  $-0.08367$ &  $-1.92138$ &  $5.16230$ &  $-1.43962$ \\
$-10.00$ & $-95.50$  &$-0.30835$ & $80.93421$ & $1.24297$ & $3.48118$ &  $1.13764$ &  $34.38667$ &  $0.45068$ &  $3.94205$ &  $-0.10264$ &  $-1.95680$ &  $5.14583$ &  $-1.44879$ \\
$-10.00$ & $-95.25$ & $-0.31660$ & $79.75581$ & $1.24499$ & $3.52980$ & $1.18787$ &  $34.52375$ &  $0.45699$ &  $3.94398$ &  $-0.10362$ &  $-1.95656$ &  $5.14707$ &  $-1.43938$ \\
$-10.00$ & $-95.00$ & $-0.34153$ & $82.27132$ & $1.26053$ & $3.62626$ & $1.19934$ &  $33.80105$ &  $0.44619$ &  $3.94591$ &  $-0.13114$ &  $-2.00462$ &  $5.12652$ &  $-1.42911$ \\
$-10.00$ & $-94.75$ & $-0.33274$ & $78.02614$ & $1.20751$ & $3.55568$ & $1.14844$ &  $34.25120$ &  $0.45313$ &  $3.94785$ &  $-0.10777$ &  $-1.96562$ &  $5.14215$ &  $-1.43711$ \\
$-10.00$ & $-94.50$ & $-0.33143$ & $80.47062$ & $1.24882$ &  $3.53934$ & $1.16409$ &  $34.25127$ &  $0.44153$ &  $3.94980$ &  $-0.10675$ &  $-1.97165$ &  $5.13509$ &  $-1.43557$\\
\enddata
\tablecomments{A complete electronic version of this table is available at: ....}
\label{tab5}
\end{deluxetable*}

\end{turnpage}
\clearpage
\global\pdfpageattr\expandafter{\the\pdfpageattr/Rotate 90}
\end{document}